\documentclass[fleqn,usenatbib,useAMS]{mnras}
\usepackage{graphicx}	% Including figure files
\usepackage{amsmath}	% Advanced maths commands
\usepackage{amssymb}	% Extra maths symbols
\usepackage{multicol}        % Multi-column entries in tables
\usepackage{bm}		% Bold maths symbols, including upright Greek
\usepackage[T1]{fontenc}
\usepackage{ae,aecompl}
\usepackage{txfonts}
\usepackage{soul, xcolor}
\setstcolor{red}

\usepackage{ulem}
\newcommand\redsout{\bgroup\markoverwith{\textcolor{red}{\rule[0.5ex]{2pt}{0.4pt}}}\ULon}

\def\nbody{$N$-body}
\def\rhl{r_{\rm hl}}
\def\rhob{r_{\rm h,ob}}
\def\feh{\rm{[Fe/H]}}
\def\ml{M/L_V}
\def\pc{{\rm pc}}
\def\kpc{{\rm kpc}}
\def\Gyr{{\rm Gyr}}
\def\rh{r_{\rm hm}}
\def\rhi{r_{\rm hm,0}}
\def\rJ{r_{\rm J}}

\def\msun{{\rm M_\odot}}
\def\Msun{{\rm M_\odot}}

\def\Rg{R_{\rm g}}
\def\Rgi{R_{\rm g,0}}
\def\Dg{D_{\rm g}}
\def\Dgcl{D_{\rm g}^{\rm cl}}
\def\Mcl{M_{\rm cl}}
\def\Mcli{M_{\rm cl,0}}

\def\kms{{\rm km}\,{\rm s}^{-1}}
\def\sigv{\sigma_{\rm v}}

\def\gtsima{$\; \buildrel > \over \sim \;$}
\def\simgt{\lower.5ex\hbox{\gtsima}}

\def\nbsix{{\sc nbody6}}
\def\nbsixdf{{\sc nbody6df}}

\defcitealias{Read2016a}{R16}
\defcitealias{Petts2016}{P16}
\title[A dark matter core in  Eridanus II]{Probing dark matter with star clusters: a dark matter core in the ultra-faint dwarf Eridanus II} 
\author[Contenta et al.]{Filippo Contenta$^1$\thanks{Contact e-mail: \href{mailto:f.contenta@surrey.ac.uk}{f.contenta@surrey.ac.uk}}, Eduardo Balbinot$^1$, James A. Petts$^1$,  Justin I. Read$^1$, Mark Gieles$^1$, \newauthor Michelle L. M. Collins$^1$, Jorge Pe\~{n}arrubia$^2$, Maxime Delorme$^1$, Alessia Gualandris$^1$\\
 $^1$ Department of Physics, University of Surrey, Guildford GU2 7XH, UK\\
 $^2$ Institute for Astronomy, University of Edinburgh, 
 Royal Observatory, Blackford Hill, Edinburgh EH9 3HJ, UK}

\date{Accepted 2018 February 13. Received 2018 February 13; in original form 2017 May 4}
\pubyear{2018}

\begin{document}
\label{firstpage}
\pagerange{\pageref{firstpage}--\pageref{lastpage}}

\maketitle

\begin{abstract}
\noindent
We present a new technique to probe the central dark matter (DM) density profile of galaxies that harnesses both the survival and  observed properties of star clusters. As a first application, we apply our method to the `ultra-faint' dwarf Eridanus II (Eri~II) that has a lone star cluster $\sim 45$\,pc from its centre. Using a grid of collisional $N$-body simulations, incorporating the effects of stellar evolution, external tides and dynamical friction, we show that a DM core for Eri~II naturally reproduces the size 
and the projected position of its star cluster. By contrast, a dense cusped galaxy requires the cluster to lie implausibly far from the centre of Eri~II \mbox{($>1\,\kpc$)}, with a high inclination orbit that must be observed at a particular orbital phase. Our results, therefore, favour a dark matter core. This implies that either a cold DM cusp was `heated up' at the centre of Eri II by bursty star formation, or we are seeing an evidence for physics beyond cold DM.
\end{abstract}


\section{Introduction}\label{sec:intro}
The $\Lambda$ `Cold Dark Matter' ($\Lambda$CDM) model gives a remarkable match to the growth of structure on large scales in the Universe \citep[e.g.][]{2002PhRvD..66j3508T,2014A&A...571A..16P}. Yet on smaller scales, inside galaxy groups and galaxies, there have been long-standing tensions \citep[e.g.][]{1999ApJ...522...82K,1999ApJ...524L..19M}. Key amongst these is the `cusp-core' problem. Pure dark matter (DM) simulations of structure formation in a $\Lambda$CDM cosmology predict that galaxies should reside within dense central DM cusps with density $\rho \propto r^{-1}$ \citep*[e.g.][]{1991ApJ...378..496D,Navarro1996} whereas observations of the rotation curves of dwarf galaxies have long favoured constant density cores \citep[e.g.][]{1994ApJ...427L...1F,1994Natur.370..629M,Read2017}. This may owe to physics beyond CDM, for example self-interacting DM (SIDM; e.g. \citealt{2000PhRvL..84.3760S,2016PhRvL.116d1302K}), wave-like DM \citep[e.g.][]{2014PhRvL.113z1302S} or ultra-light axions \citep[e.g.][]{Gonzales2016}. However, all of these small-scale tensions with $\Lambda$CDM arise when comparing models devoid of `baryons' (stars and gas) with real galaxies in the Universe. There is mounting evidence that bursty star formation during galaxy formation can `heat-up' DM, transforming a DM cusp to a core (e.g. \citealt*{1996MNRAS.283L..72N}; \citealt{2005MNRAS.356..107R,2012MNRAS.421.3464P,2014Natur.506..171P,2015MNRAS.451.1366P}). The latest simulations, that reach a mass and spatial resolution sufficient to resolve the multiphase interstellar medium, find that DM cores, of approximately the half stellar mass radius in size ($R_{1/2}$), form slowly over a Hubble time \citep{2008Sci...319..174M,2010Natur.463..203G,2012MNRAS.421.3464P,2013MNRAS.429.3068T,2014ApJ...789L..17M,2014MNRAS.441.2986D,2015MNRAS.454.2092O,2016MNRAS.456.3542T,Read2016a,2017arXiv170506286M}. 

Although the above simulations agree on the size and formation timescale of DM cores, there remains some disagreement over the DM halo mass at which cusp-core transformations become inefficient, $M_{200} \equiv M_{\rm pristine}$\footnote{$M_{200}$ is the virial mass. For satellite galaxies, we define this pre-infall.}. As pointed out by \citet{2012ApJ...759L..42P}, for a fixed $M_{200}$ if too few stars form then there will no longer be enough integrated supernova energy to unbind the DM cusp. Depending on the numerical scheme employed, $M_{\rm pristine}$ has been reported to be as high as $M_{\rm pristine} \sim 10^{10}$\,M$_\odot$ \citep[e.g.][]{2015MNRAS.454.2981C,2016MNRAS.456.3542T} and as low $M_{\rm pristine} \sim 10^8$\,M$_\odot$ \citep[][hearafter \citetalias{Read2016a}]{Read2016a}, where the spread owes primarily to different star formation efficiencies in low mass halos (see \S\ref{subsec:DM_profile} and Fig.~\ref{fig:ms_mh}).

The above motivates measuring the central DM density of the very faintest galaxies in the Universe. With little star formation, these may be expected to retain their `pristine' DM cusps (e.g. \citetalias{Read2016a}). There is no shortage of such faint dwarf galaxies orbiting the Milky Way, Andromeda and nearby systems \citep[e.g.][]{2007ApJ...654..897B,2014ApJ...783....7C,Bechtol2015,2015ApJ...812L..13S}. However, most of these are devoid of gas and so the kinematics of their stars must be used to probe their DM halos. This is challenging because of a strong degeneracy between their DM density profiles and the orbit distribution of their stars \citep[e.g.][]{1990AJ.....99.1548M,2009MNRAS.393L..50E,2017MNRAS.471.4541R}. For the brighter Milky Way dwarfs, this degeneracy can be broken by using metallicity or colour to split the stars into distinct components with different scale lengths (e.g. \citealt{2008ApJ...681L..13B,2011ApJ...742...20W} and \citealt{2012ApJ...754L..39A}, but see \citealt{2013A&A...558A..35B} and \citealt{2014MNRAS.441.1584R}). However, for the fainter dwarfs there are too few stars to obtain strong constraints \citep{2017MNRAS.471.4541R}.

An alternative method for probing the central density of dwarf galaxies was proposed by \citet{1998MNRAS.297..517H}, \citet{Goerdt2006} and \citet{2006MNRAS.370.1829S}. They showed that the globular clusters (GCs) in the dwarf spheroidal galaxy Fornax would rapidly sink to the centre by dynamical friction if Fornax has a steep DM cusp. By contrast, in a constant density core, dynamical friction is suppressed \citep{Read2006a,Inoue2009,Inoue2011,Petts2015,Petts2016}, allowing Fornax's GCs to survive through to the present day. This `timing argument' was refined by \citet{ColeD2012} who used 2800 $N$-body simulations of Fornax's GC system to show that a core is favoured over a cusp, in excellent agreement with split-population modelling of Fornax's stars \citep[e.g.][]{2011ApJ...742...20W}. (Such survival arguments were extended by \citet{2009MNRAS.399.1275P} to the GCs associated with the Sagittarius dwarf.) Although it is likely that Fornax has a DM core, its stellar mass ($M_* \sim 4 \times 10^7$\,M$_\odot$; \citealt{2012A&A...544A..73D}) is large enough for bursty star formation to drive complete cusp-core transformations (\citealt{2012ApJ...759L..42P}; \citetalias{Read2016a}). Thus, Fornax's core yields inconclusive constraints on the nature of DM.

In this paper, we develop a new method for probing the central DM density of dwarf galaxies that harnesses both the survival and present-day properties of star clusters. Star clusters are dense stellar systems that slowly expand due to two-body relaxation \citep{H65, 2010MNRAS.408L..16G}. In a tidal field, high-energy stars are pushed over the cluster's tidal boundary, slowing down the expansion. Eventually, the cluster's half stellar mass radius becomes a constant fraction of the tidal radius and, from that moment on, the cluster evolves approximately at a constant density set by the tidal field \citep*{H61, 2011MNRAS.413.2509G}. Thus, the observed surface density of low-mass GCs (i.e. those that have undergone sufficient relaxation) can be used as probes of the host galaxy's tidal field and, therefore, its density distribution \citep{1983AJ.....88..338I}. This allows us to probe the DM distribution in any dwarf galaxy with low-mass star clusters, including those with a much lower stellar mass than Fornax.  This is the key idea that we exploit in this work\footnote{Note that the cluster's stellar kinematics are also affected by tides, making them additional probes of the properties of the galactic tidal field \citep[e.g.][]{2010MNRAS.407.2241K, 2017MNRAS.466.3937C}.}.

To model star clusters sinking in the potential of a host dwarf galaxy, we make use of the semi-analytic dynamical friction model from \citet{Petts2016} (hereafter \citetalias{Petts2016}), implemented in the direct-summation code \nbsix\ \citep{2003gnbs.book.....A}. This allows us to model the survival of star clusters, similarly to \citet{ColeD2012}, but with a complete $N$-body model of the star cluster itself, including two-body effects, binary formation and evolution and stellar evolution. By comparing a large grid of such models with observational data, we are able to constrain the DM density of dwarf galaxies that host low mass GCs, independently of timing arguments or stellar kinematic measurements.

As a first application, we apply our method to the ultra-faint dwarf galaxy Eridanus II (Eri~II) that was recently discovered by the Dark Energy Survey (DES; \citealt{Bechtol2015,Koposov2015}). Eri~II is situated $366\,\kpc$ from the Sun, at the edge of the MW, with $M_V=-7.1$, a half-light radius of $R_{1/2} = 2.31'$, and an ellipticity of $0.48$. Eri~II appears to show an extended star formation history, but follow-up observations are needed to confirm this. \citet{Koposov2015} and \citet{Crnojevic2016} found that Eri~II has a lone star cluster at a projected distance $\sim45\,\pc$ from Eri~II's centre, with $M_V=-3.5$ and a half-light radius of $13\,\pc$ \citep[see][Table 1]{Crnojevic2016}. Compared to the MW's star clusters \citep[][2010 edition]{Harris1996}, Eri~II's star cluster appears faint and extended, contributing just $\sim4\%$ of Eri~II's total luminosity.

This paper is organised as follows. In \S\ref{sec:method}, we describe our method for probing the central DM density of dwarf galaxies using star clusters, and we motivate our priors for modelling Eri~II. In \S\ref{sec:results}, we present our main findings. In \S\ref{sec:discussion}, we discuss the implications of our results for galaxy formation and the nature of DM. Finally, in \S\ref{sec:conclusions}, we present our conclusions.

\section{Method}\label{sec:method}

\subsection{A new method for measuring the inner DM density of dwarf galaxies}\label{subsec:methodoverview}

We model the evolution of star clusters orbiting within a host dwarf galaxy using \nbsixdf\ (\citetalias{Petts2016}). This is a publicly available\footnote{ \url{http://github.com/JamesAPetts/NBODY6df}.} adaptation of \nbsix, which is a fourth-order Hermite integrator with an \citet{1973JCoPh..12..389A}
neighbour scheme \citep{1992PASJ...44..141M, 1999PASP..111.1333A, 2003gnbs.book.....A}, 
and force calculations that are accelerated by Graphics Processing
Units \citep[GPUs,][]{2012MNRAS.424..545N}. \nbsix\ contains metallicity dependent prescriptions for the evolution of individual stars and binary stars \citep{2000MNRAS.315..543H, 2002MNRAS.329..897H} which we use in our simulations here.

In \nbsixdf, we model the host dwarf galaxy as a static, analytic potential. Dynamical friction is then applied to star cluster members using the semi-analytic model described in \citetalias{Petts2016} \citep[see also][]{Petts2015}. The \citetalias{Petts2016} model has been extensively tested against full $N$-body simulations of dynamical friction in both cored and cusped background potentials, giving an excellent description of the orbital decay in both cases. In particular, it is able to reproduce the `core-stalling' behaviour, whereby dynamical friction is  suppressed inside constant density cores (\citealt{Goerdt2006,Read2006a,Inoue2009,Inoue2011} and see \citetalias{Petts2016} for further details).

We set up a grid of 200 \nbsixdf\ simulations, varying the density profile (cusped or cored) and the initial orbit and properties of the star cluster. Comparing this grid with observations, we determine the most likely mass distribution for Eri~II, and the initial properties of its star cluster. (Eri~II is dominated at all radii by its dark matter halo (see \S\ref{subsec:DM_profile}) and so its total mass distribution directly provides us with its DM density profile.) In addition, we run a further 26 simulations to determine how our results depend on the mass, concentration and inner logarithmic slope of Eri~II's dark matter halo (\S\ref{subsec:diff_M200_c200}). We also test whether Eri~II's cluster could form and survive in the very centre of Eri~II (\S\ref{subsec:nuclear_cluster}).

\subsection{The DM halo of Eri~II}
\label{subsec:DM_profile}

We model the DM halo of Eri~II using the coreNFW profile from \citetalias{Read2016a}. This is described by a mass $M_{200}$ and concentration parameter $c_{200}$, identical to those used for the cusped Navarro-Frenk-White (NFW) profile \citep*{Navarro1996}. However, it allows also for a central DM core. By default, this has a size set by the projected half light radius of the stars $R_{1/2}$, which for Eri~II is $R_{1/2} = 0.28$\,kpc\footnote{
The latest estimates of $R_{1/2}$ are slightly lower than the value we have assumed here, $R_{1/2} = 246\pm13\,$pc (Denija Crnojevi\'{c}, private communication), though within the $2\sigma$ uncertainties. However, as we show in \S\ref{subsec:diff_M200_c200}, we are not very sensitive to $M_{200}$, $c_{200}$ or the DM core size. As such, the newer value for $R_{1/2}$ will not affect our results.} \citep{Crnojevic2016}. The power-law slope of the core is set by $n$, where $n=1$ produces a flat dark matter core, whereas $n=0$ returns the fully cusped NFW profile\footnote{In \citetalias{Read2016a}, $n$ was parameterised by the total star formation time. However, since this is poorly determined for Eri~II, we consider here just a range of values for $n$. We discuss this further in \S\ref{sec:discussion}.}.

Eri~II's stellar population may be expected to be like other similar stellar mass dwarf galaxies (e.g. Bootes I and Ursa Major I; see \citealt{McConnachie2012}) which have predominantly old-age stars. \citet{Santana2013} showed that the apparent intermediate-age population in these galaxies is likely due to the presence of blue straggler stars. However, they could not rule out the presence of an intermediate-age population of up to 3\,Gyr old. Recent HST data (propID 14234; subject of a future publication), has confirmed that Eri~II's stellar population is similar to those studied by \citet{Santana2013}, hence favoring an older population. In this work, we choose to be conservative and assume that the cluster is older than $5\,$Gyr.

\begin{figure}
\center
\includegraphics[width=0.48\textwidth]{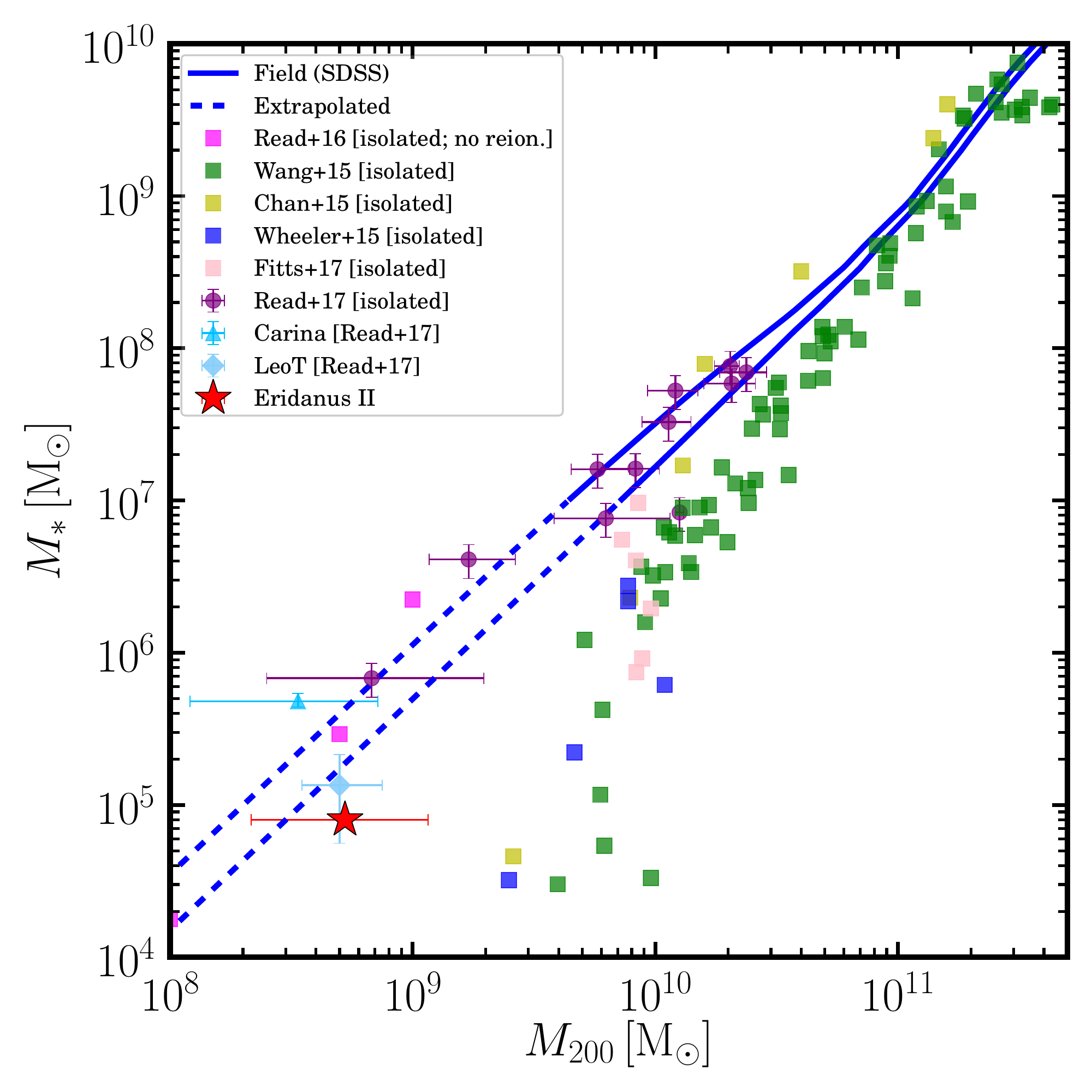}
\caption{The stellar mass-halo mass relation of isolated dwarf galaxies. The symbols correspond to the observational data and the squares to the results from \nbody\ simulations. The blue solid lines show the results from abundance matching in $\Lambda$CDM using the SDSS field stellar mass function (where the lines are dashed, the results are extrapolated). Eri~II is marked by the red star. It  appears to be consistent with a `failed' Leo T, inhabiting a similar DM halo but having its star formation shut down earlier, lowering its $M_*$ for the same pre-infall $M_{200}$. 
}\label{fig:ms_mh}
\end{figure}

To obtain an estimate of the (pre-infall) halo mass, $M_{200}$, for Eri~II we use the recent measurement of its mass within the 3D half light radius $M_{1/2} = 1.2_{-0.3}^{+0.4} \times {10}^7\,\Msun$, derived from stellar kinematics by \citet{2017ApJ...838....8L}. We turn this into an $M_{200}$ by fitting an NFW profile to $M_{1/2}$ using the $M_{200}-c_{200}$ relation from \citet{2007MNRAS.378...55M}, finding $M_{200} = 4.7^{+6.9} _{-2.6}\times 10^8\,\Msun$.

To test if the above value for $M_{200}$ is reasonable, in Fig.~\ref{fig:ms_mh} we compare the stellar mass ($M_* \simeq 8 \times 10^4\,\Msun$; \citealt{Bechtol2015}) and $M_{200}$ for Eri~II with measurements for other nearby dwarfs; Eri~II is marked by the red star. The purple circles show the $M_*-M_{200}$ relation for isolated gas-rich dwarfs from \citet{Read2017}. The dark cyan triangle shows a measurement for the Carina dwarf spheroidal galaxy from \citet{2015NatCo...6E7599U}. The light cyan diamond shows an estimate for the isolated dIrr Leo T from \citet{Read2017}. The blue solid and dashed lines show the $M_*-M_{200}$ relation derived by \citet{Read2017} from abundance matching in $\Lambda$CDM using the SDSS field stellar mass function (the dashed lines show where this is extrapolated). The remaining data points show the latest results from a range of simulations of isolated dwarfs taken from the literature: \citetalias{Read2016a} (magenta); \citet{2015MNRAS.454...83W} (green); \citet{2015MNRAS.454.2981C} (yellow); \citet{2015MNRAS.453.1305W} (blue); and \citet{2017MNRAS.471.3547F} (pink). As can be seen, there is a clear discrepancy between most simulations and the data below $M_{200} \sim 10^{10}$\,M$_\odot$ that remains to be understood \citep{2018MNRAS.473.2060J}. For our paper here, however, this plot demonstrates that our derived $M_{200}$ for Eri~II is in good agreement with estimates for other galaxies of a similar stellar mass. Eri~II is consistent with a `failed' Leo T, inhabiting a similar DM halo but having its star formation shut down earlier, lowering its $M_*$ for the same pre-infall $M_{200}$. This is further evidenced by the lack of detected HI gas, or recent star formation, in Eri~II \citep{Crnojevic2016}.

\begin{figure*}
\center
 \includegraphics[width=.48\textwidth]{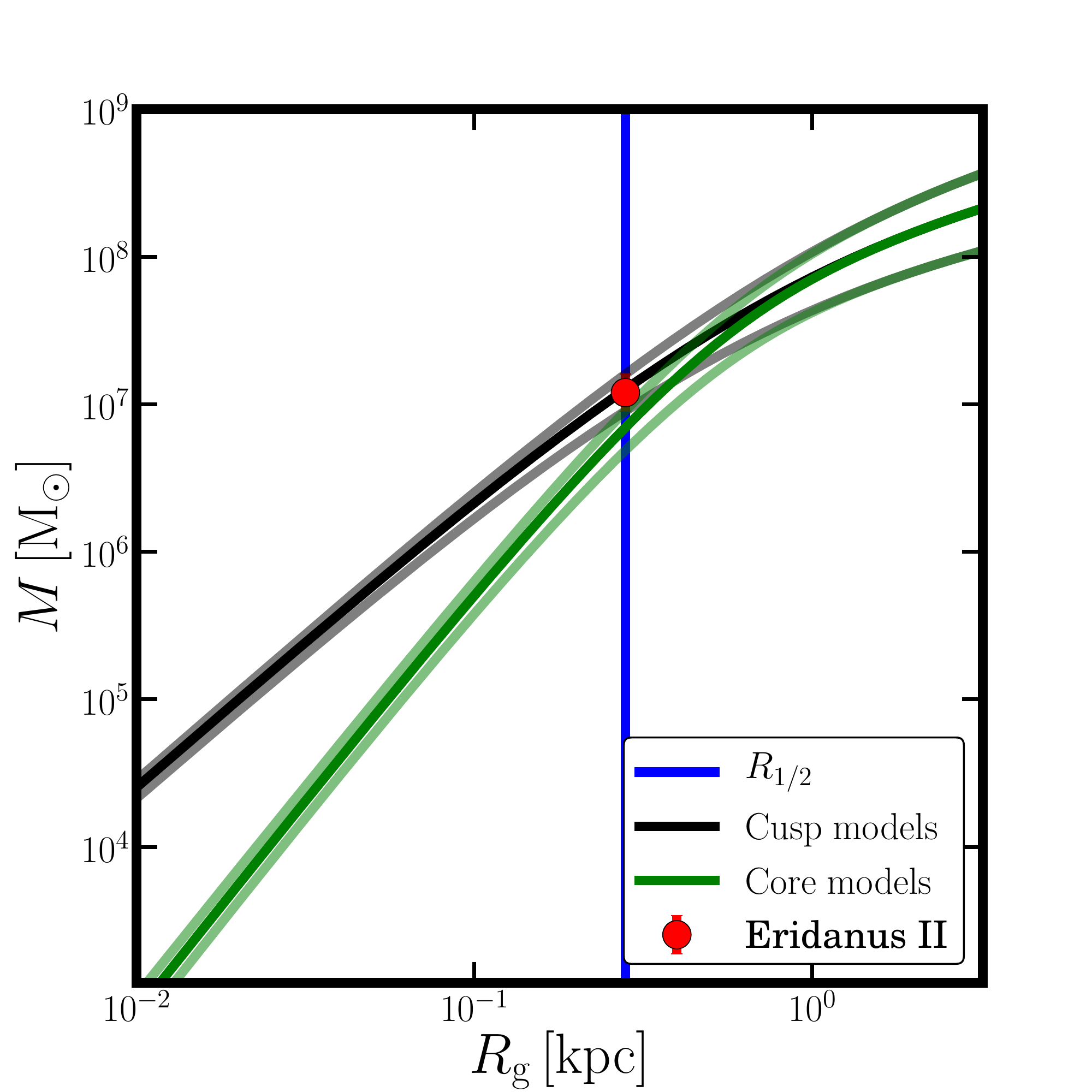}%
 \includegraphics[width=.48\textwidth]{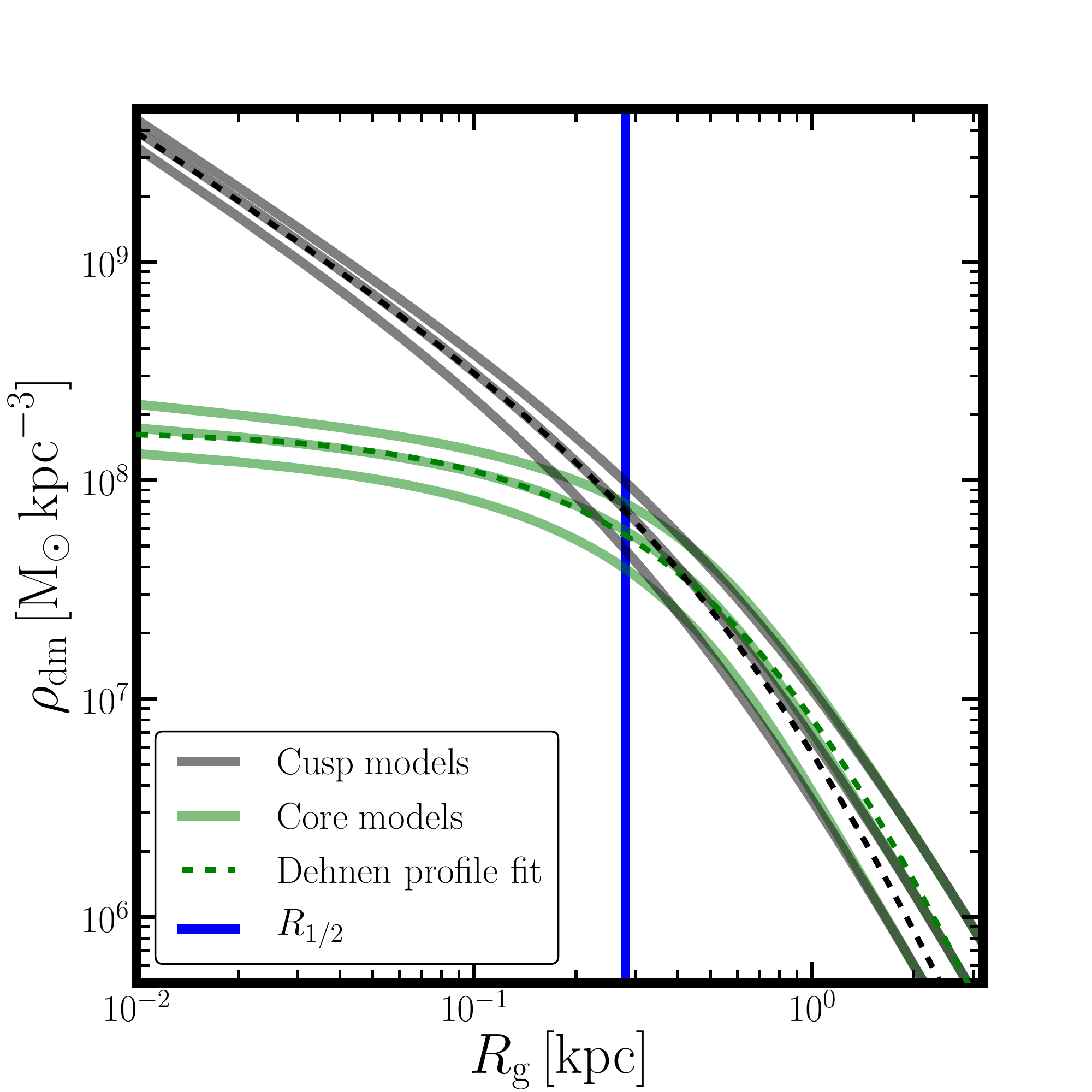}
\caption{DM halo models for Eri~II, chosen as described in \S\ref{subsec:DM_profile}. The left panel shows the cumulative mass profiles; the right panel the logarithmic density profiles. The mass within the projected stellar half light radius $M_{1/2}$ is marked by the red data point with error bars on panel a) (taken from \citealt{2017ApJ...838....8L}). The dashed black and green lines show the cusped and cored models explored in this work, respectively. These are the best-fit Dehnen profile models to the dwarf galaxy models marked by the solid black and green lines, respectively. For our fiducial grid of simulations, we assume Eri~II inhabits DM halo with a virial mass of $M_{200} = 5 \times 10^8$\,M$_\odot$. Models with a halo mass at the 68\% upper and lower bound of Eri-II's stellar kinematics are marked by the upper and lower gray and green lines. For the core profiles, we used the same halo mass as the cusp profiles with $n=0.9$ (corresponding to a Dehnen model with $\gamma=0$; dashed green line). The vertical blue line on both panels marks the projected half light radius of Eri~II, $R_{1/2}$. We explore the effect of varying $M_{200}$, $c_{200}$ and the inner logarithmic cusp slope of Eri~II's DM halo in \S\ref{subsec:diff_M200_c200}.}
\label{fig:EriII_profile}
\end{figure*}

In Fig.~\ref{fig:EriII_profile}, we show the cumulative mass profiles (left panel) and DM density profiles (right panel) for Eri~II that we assume in our fiducial grid of simulations (we explore different halo masses, concentrations and logarithmic cusp slopes in \S\ref{subsec:diff_M200_c200}). The grey lines show the cusped model, the green lines show the cored model. The middle of the three lines shows $M_{200} = 5 \times 10^8\,\Msun$ that we assume from here on. The top and bottom lines show the upper and lower boundary of $M_{200}$ estimated from the kinematic measurements \citep{2017ApJ...838....8L}. The projected half light radius of the stars, $R_{1/2}$, is marked by the vertical blue line. On the left panel, the measurement of $M_{1/2}$ for Eri~II from \citet{2017ApJ...838....8L} is marked by the red data point. As can be seen, due to the $M_{200}-c_{200}$ relation, changing $M_{200}$ produces only a small effect on the DM density within $R_{1/2}$. Thus, our method will not be very sensitive to $M_{200}$ (we will verify this expectation in \S\ref{subsec:diff_M200_c200}). However, cusped and cored models look very different within $R_{1/2}$ and this is what we aim to probe in this work.

Finally, the latest version of \nbsixdf\ only supports a background DM density profile modelled by Dehnen spheres \citep{Dehnen1993}:

\begin{equation}
\rho(r) = \frac{M_0(3-\gamma)}{4\pi r_0^3}\left(\frac{r}{r_0}\right)^{-\gamma}\left(1+\frac{r}{r_0}\right)^{\gamma-4},
\label{eqn:dehnen}
\end{equation}
where $M_0$ and $r_0$ are the mass and scale length, respectively, and $-\gamma$ is the logarithmic slope of the inner density profile.

Thus, to obtain DM profiles suitable for \nbsixdf, we fit the above Dehnen profile to our coreNFW density profiles. These fits are shown by the dashed lines in the right panel of Fig.~\ref{fig:EriII_profile}. As can be seen, inside $R_{1/2}$ (our region of interest), these fits are excellent. 
Our best-fit parameters for the cored ($\gamma = 0$, which corresponds to $n=0.9$ for our coreNFW profile) and cusped ($\gamma = 1$, which corresponds to $n=0$) models were: $M_0 = 4.79 \times 10^8\,\Msun$ and $r_0 = 0.877$\,kpc and $M_0 = 2.94 \times 10^8\,\Msun$ and $r_0 = 1.078$\,kpc, respectively.

\subsection{Eri~II's star cluster}\label{subsec:EriII_starcluster}
We model the initial conditions of Eri~II's star cluster as a Plummer sphere \citep{Plummer1911} with a Kroupa IMF \citep{Kroupa2001}, sampling stars with masses between $0.1\,\msun$ and $100\,\msun$ and assuming a metallicity of $Z=0.0008$ (corresponding to $\feh\simeq-1.5$). We assumed a range of initial masses $\Mcli$ and half-mass radii $\rhi$ for the cluster to explore how its initial properties impact its final state. 

\subsection{Exploring parameter space}\label{subsec:param_space}
To explore the parameter space, we ran 200 simulations, 100 for each galaxy model (core and cusp). We varied $\rhi$, $\Mcli$ and the initial galactocentric distance ($\Rgi$) of the cluster.
We allowed the cluster to have a $\rhi$ of 1, 5, 10, 15 and 20$\,\pc$; a $\Mcli$ of approximately 13,000, 19,000, 25,000 and 32,000$\,\Msun$; and $\Rgi$ of 0.14, 0.28, 0.56, 1.12 and $2.8\,\kpc$. The $\rhi$ range is based on what is found for young massive clusters \citep{2010ARA&A..48..431P}. The minimum $\Mcli$ is chosen such that after stellar mass loss the mass is always above the mass of the cluster. The maximum mass was chosen such that less than $20\%$ of all stars in the entire galaxy originated from the star cluster, which is a reasonable upper limit \citep{2012A&A...544L..14L,2014A&A...565A..98L}.

For all clusters we adopted circular orbits. This favours the survival of clusters in cusped profiles because eccentric orbits reach closer to the centre of the galaxy where clusters are less likely to survive. (Our assumption that the host dwarf galaxy has a spherical potential similarly favours a cusped profile because in triaxial models there are no circular orbits and only the more damaging radial orbits are allowed.)

\citet{2017MNRAS.466.1741C} show how the observations of faint star clusters -- like Eri~II's star cluster -- can be affected by primordial binaries and the retention fraction of black holes, together with observational biases. In our simulations we did not vary these aspects, which could be degenerate with the initial conditions of the clusters, nor did we vary the initial density profile of the clusters.  

\begin{figure}
\center
\includegraphics[width=0.48\textwidth]{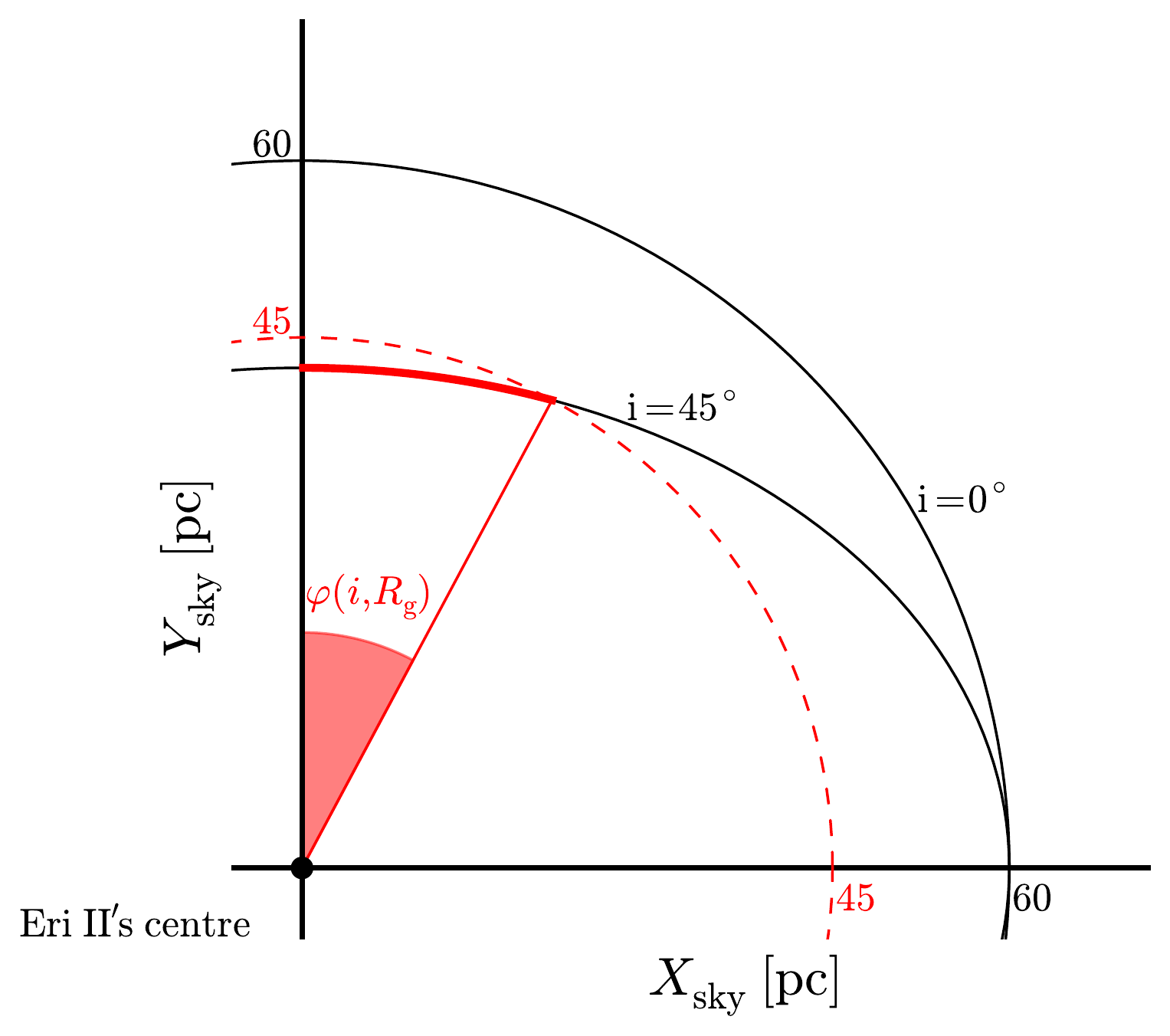}
\caption{Schematic representation of two projected orbits with different inclinations of their orbital plane $i$ (solid black lines). Eri~II's cluster is observed to be $45\,\pc$ from the centre of Eri~II in projection (dashed red line). We consider that a cluster can be observed as Eri~II's star cluster if its orbit is within $45\,\pc$ (solid red line). For a given $i$, a larger distance from the centre results in a smaller $\varphi(i,\Rg)$  (see equation~\ref{eq:varphi}). Therefore, clusters that orbit in the outskirts of Eri~II are unlikely to be observed near the centre.
} \label{fig:schematic_orbit}
\end{figure}

Eri~II's star cluster is observed at a projected distance $\Dgcl=45\,\pc$ from the centre of Eri~II. Thus, we also need to take into account the probability for the cluster to be observed at that radius in the total likelihood. We estimate the probability $P(\Dg<\Dgcl|\Rg)$ to observe a cluster (on a circular orbit) within $\Dgcl$ for a given $\Rg$, assuming a random inclination of the orbital plane with respect to the observer. To compute $P(\Dg<\Dgcl|\Rg)$, firstly we estimate the angle $\varphi(i,\Rg)$, which defines the angle in which the cluster is observed to be within $\Dgcl$ during 1/4 of an orbit (see  Fig.~\ref{fig:schematic_orbit}), where $i\in[0,\pi/2]$ is the angle between the pole of the orbit and the line of sight. For circular orbits, the angle $\varphi(i,\Rg)$ is given by

\begin{equation}\label{eq:varphi}
\varphi(i,\Rg) = \begin{cases}
\displaystyle
\frac{\pi}{2} - \arcsin \left( \frac{\sqrt[]{1-\Dgcl/\Rg}}{\sin i}\right),&\Rg>\Dgcl,\\
\displaystyle\frac{\pi}{2},&\Rg<\Dgcl.
\end{cases}
\end{equation}
Secondly, we integrate $\varphi(i,\Rg)$ with respect to $\cos i$, because for random inclinations of the orbital plane, $\cos i$ is uniformly distributed. We then divide by a normalization angle, $\pi/2$, because $\varphi(i, \Rg)$ considers only 1/4 of an orbit, see Fig.~\ref{fig:schematic_orbit} and we obtain

\begin{equation}\label{eq:P}
P(\Dg<\Dgcl|\Rg)= \frac{2}{\pi} \int^1_0 \varphi(i,\Rg) \ {\rm d}\cos i.
\end{equation}
By definition, $0\le P(\Dg<\Dgcl|\Rg)\le1$. 
 
To compare the \nbody\ simulations with the observational data, we assumed a stellar mass-to-light ratio of $\ml=2$, appropriate for old, metal-poor stellar populations \citep[e.g.][]{McLaughlin2005}, obtaining $M_{\rm ob}=4.3\times10^3\,\Msun$. We multiply the observed half-light radius ($\rhl$) by  $4/3$ to  correct  for projection effects \citep{Spitzer1987}, to get an estimate for the 3D half-mass radius $\rhob=17.3\,\pc$\footnote{Assuming that light traces mass, which is not necessarily true if the cluster is mass segregated.} of Eri~II's star cluster.
To find the model that best fit the observational data, we maximise the likelihood for the fitting parameters ($\rhi$, $\Mcli$, and $\Rgi$). The log-likelihood function is:
\begin{equation}\label{eq:L}
\ln\mathcal{L} =  -\frac{\left(\rhob - \rh\right)^2}{2\sigma_{\rm r}^2} -\frac{\log^2\,(M_{\rm ob}/\Mcl)}{2\sigma_{\log(M)}^2} + \ln\,P(\Dg<\Dgcl|\Rg),  
\end{equation}
where $\sigma_{\rm r}=1.33\,\pc$ and $\sigma_{\log(M)}=0.24$\footnote{$\sigma_{\rm r}$ is the uncertainty on the 3D half-mass radius; whereas $\sigma_{\log(M)}$ is estimated assuming $\ml=2$.} are the uncertainties derived from the observation \citep{Crnojevic2016}. The last term in the equation above is given by equation~(\ref{eq:P}) and acts as a prior to our likelihood given that it penalises models that are less likely to be observed simply due to geometrical constraints. Additional free parameters, such as the time when the cluster appears at $45\,$pc, could be included, however the age and the orbit of the cluster are unknown. Whereas, the time span when the cluster reproduce the observations would favour the clusters in the cored galaxy (see \S~\ref{sec:best_fit_models}), without changing the main results. Therefore we took the simplest approach including only the geometric effects.

By computing $\rh$, $\Mcl$ (defined as the sum of the mass of all stars within the tidal radius of the cluster) and $\Rg$, we calculate the likelihood (equation~\ref{eq:L}) for each output time of the simulation.

\subsection{Estimation of the number density profile}
To study the structural properties of the clusters in the \nbody\ simulations, we used a maximum likelihood fit following the procedure described in \citet{Martin2008}. We model the stellar distribution of the clusters using 2D elliptical Plummer and spherical, single-component, King models \citep{1966AJ.....71...64K}. We fit the models with a Monte Carlo Markov Chain (MCMC) method \citep[{\sc emcee} code,][]{Foreman2013} to optimize the following parameters: projected half-number radius, ellipticity, position angle, and surface density background for the Plummer models; and half-number radius, central dimensionless potential and background surface density for the King models\footnote{We used the {\sc limepy} code \citep{Gieles&Zocchi2015} to compute the projected density profiles of King models (\url{https://github.com/mgieles/limepy}).}.

\subsection{Qualitative estimation of the size of Eri~II's star cluster}
\begin{figure}
\center
\includegraphics[width=0.48\textwidth]{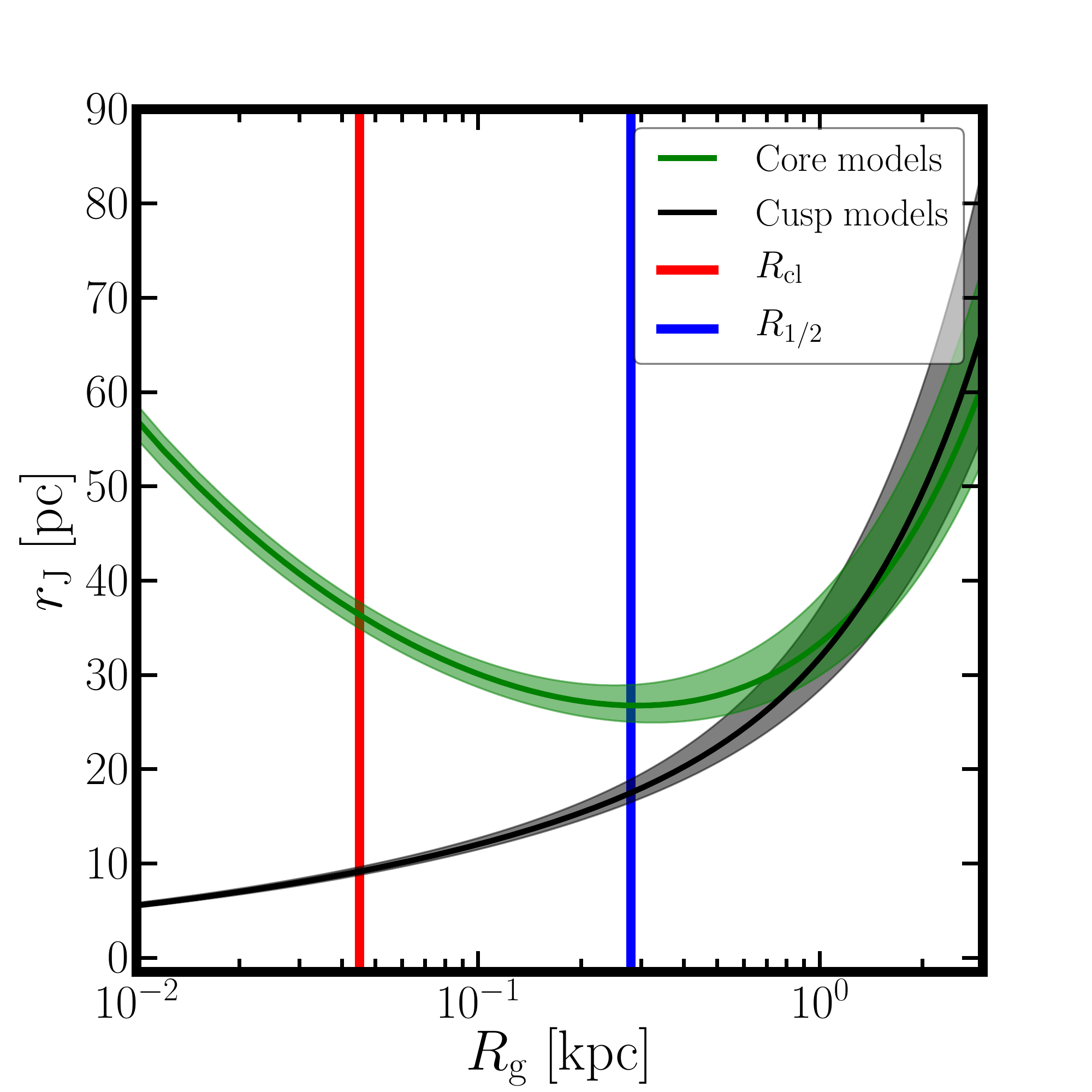}
\caption{Estimation of the tidal radius ($\rJ$) of Eri~II's star cluster for a cored (green lines) and cusped (black lines) dark matter model for Eri~II. The red and blue lines mark the observed position of the cluster and the half-light radius of the Eri~II, respectively. We assume that the cluster has a fixed mass and is on a circular orbit. Notice that the cored models allow for the cluster to have much a larger size in the inner part of the galaxy than the cusped models.}
\label{fig:cluster_rJ}
\end{figure}

We can estimate the maximum radius that a star cluster can have, which corresponds to the situation in which the cluster fills the Roche volume. As described by \citet{H61}, a star cluster in a tidal field, evolves at a constant mean density once it fills the Roche volume, which means that the ratio between the $\rh$ and the Jacobi radius ($\rJ$) is constant: $\rh/\rJ \simeq0.15$.
 The Jacobi radius is defined in \citet{1962AJ.....67..471K} as:

\begin{equation}\label{eq:King formula}
\rJ =\left(\frac{G \Mcl}{{\Omega}^2-\frac{\partial^2\phi}{\partial \Rg^2}}\right)^{\frac{1}{3}}, 
\end{equation}

\noindent where 
$\Omega$ is the angular velocity of the cluster around the Galaxy centre, 
 $\phi$ is the potential of the Galaxy and  
 $G$ is the gravitational constant. 
  In Fig.~\ref{fig:cluster_rJ}, we show that if we use the Dehnen models chosen in Fig.~\ref{fig:EriII_profile} and $\Mcl=M_{\rm ob}$, we can estimate the tidal radius of the cluster as a function of $\Rg$, assuming that the cluster is on a circular orbit. From this Figure, using the ratio $\rh/\rJ \simeq0.15$, we find that in a cusped galaxy at $\Rg=45\,$pc (red line), a cluster can be at most as large as $\rh \sim 1.5\,$pc, whereas in a cored galaxy it can be substantially larger ($\rh \sim 6$\,pc).
  
In practice, however, the ratio $\rh/\rJ$ depends on the galactic potential and the number of stars in the cluster. \citet{1997MNRAS.286..709G} found that depending on these, it can be as high as $\rh/\rJ = 0.4$. For this reason, we require full \nbody\ simulations to obtain quantitative constraints on the central dark matter density slope in Eri~II. Nonetheless, Figure \ref{fig:cluster_rJ} does give us the correct intuition that clusters will be larger in cored rather than cusped galaxies.
  
\section{Results}\label{sec:results}

\subsection{Cusp vs. core}\label{sec:cuspVScore}
\begin{figure*}
\center
\includegraphics[width=\textwidth]{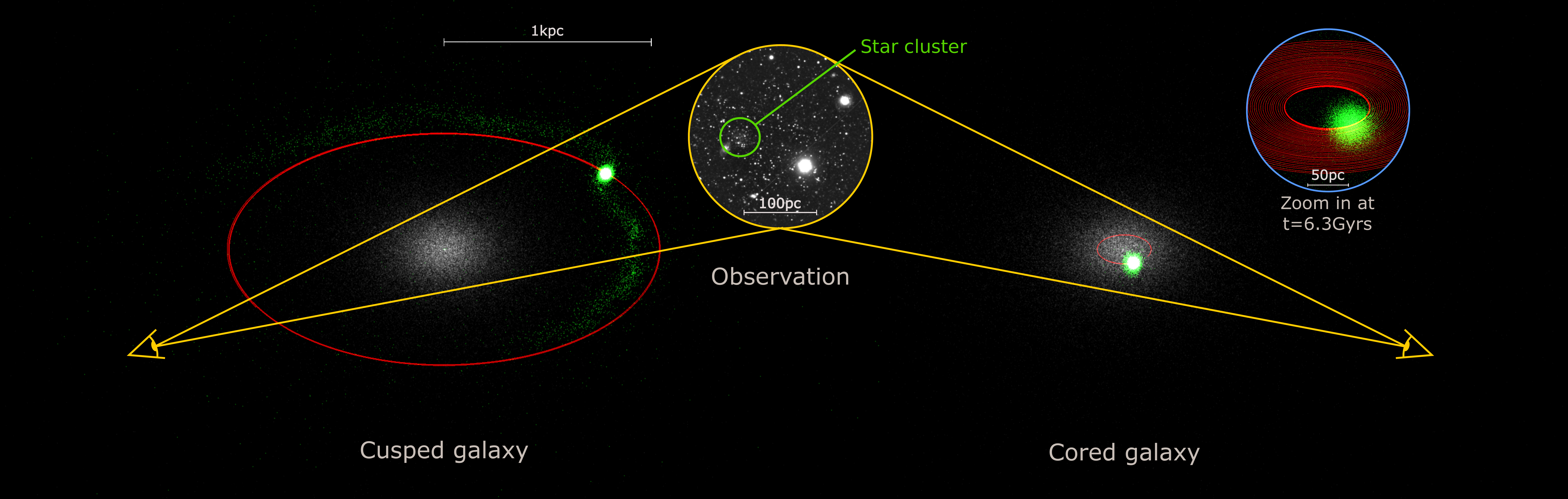}
\caption{Schematic representation of the cluster orbiting in a cusped Eri~II (on the left) and a cored Eri~II (on the right) at times when they best reproduce the observations (\citealt{Crnojevic2016}; shown in the middle). On the top right, we show a zoom-in of the simulation in the cored galaxy after 6.3 Gyr. The effect of dynamical friction and stalling can be seen in the red lines that depict the cluster's orbit. In the cusped case, the cluster can only survive in the outskirts of Eri~II, whereas in the cored case the cluster can survive in the central region where it is much more likely to match the observations.}
\label{fig:double}
\end{figure*}

The main result of our investigation is that the presence of a cored DM profile in Eri~II allows a star cluster to not only survive in the centre of the galaxy, but also to expand up to $\rh\simeq17\,\pc$ (i.e. $\rhl\simeq13\,\pc$), in excellent agreement with observations of Eri~II's lone star cluster. By contrast, a cusped DM profile gives a poorer fit overall and requires special conditions that have to be satisfied. In Fig.~\ref{fig:double}, we show a schematic representation of a simulated cluster in a cusped galaxy (on the left) and a cored galaxy (on the right) at times when they best reproduce the observations (shown in the middle). As can be seen, in the cored galaxy (right), the star cluster (green) stalls at a radius $\sim 45$\,pc from Eri~II's centre (see the zoomed image in the blue circle that shows its orbital decay and stalling in red). In this case, no special  inclination or time are required to reproduce Eri~II's star cluster. Notice also that the cluster appears visibly extended, similarly to Eri~II's cluster, and that it shows little to no tidal tails, as expected for a cluster orbiting in a constant density core \cite[e.g.][]{Petts2016}. By contrast, in the cusped case (left), the cluster must orbit much farther ($\Rg>1\,\kpc$) from Eri~II's centre in order to survive. Now its orbit will only be close enough to Eri~II in projection when the red circle lies inside the two solid yellow lines. This happens when the cluster orbits with a high inclination ($i>87.43^\circ$) of the orbital plane, and is in a particular orbital phase (that occurs for <3\% of the total orbit time). In the cusped galaxy, the cluster is denser than in the cored case and less consistent with the data for Eri~II's cluster. There are also now two visible tidal tails, as expected for a cluster orbiting in a cusped background.

\begin{figure*}
\center
\includegraphics[width=.32\textwidth]{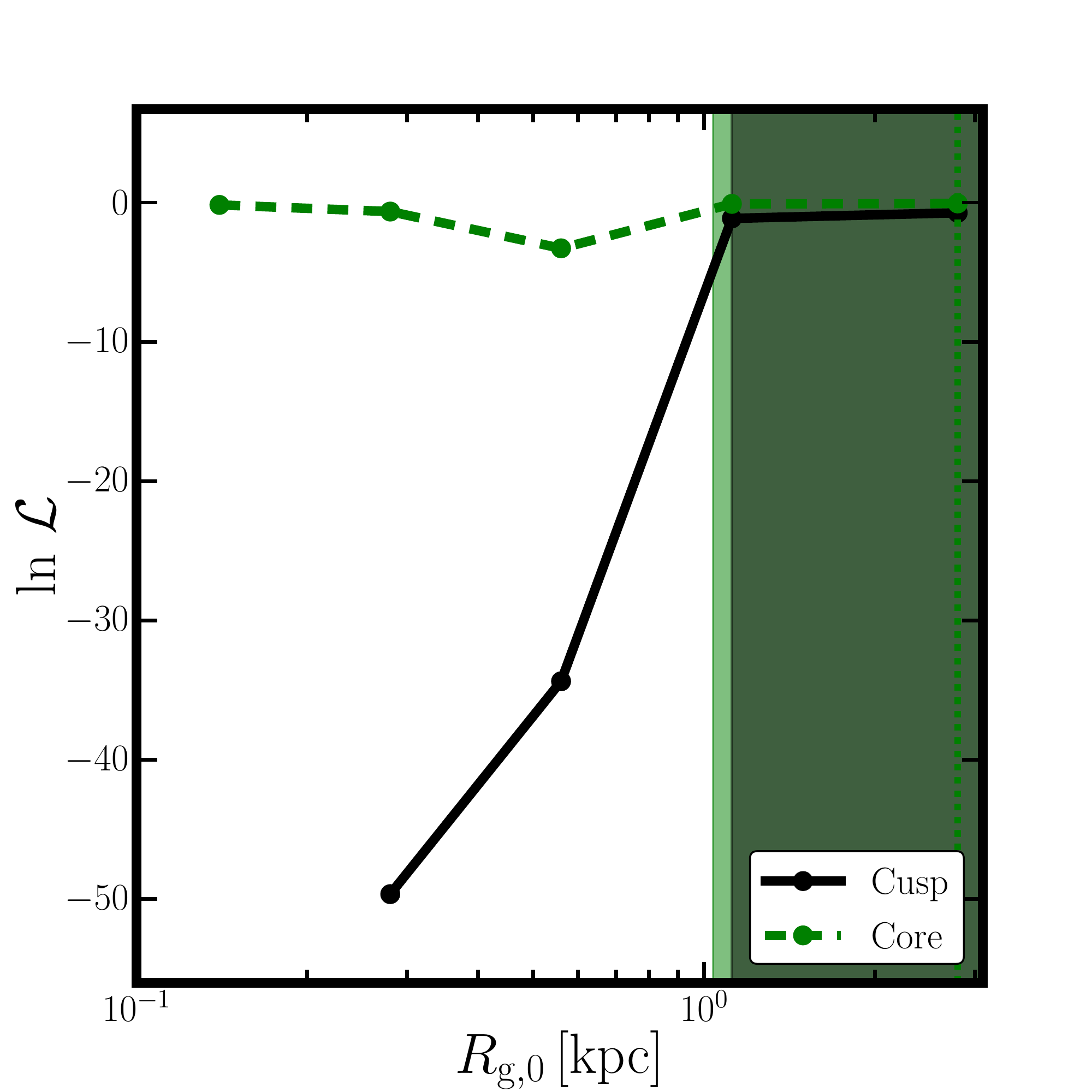}
\includegraphics[width=.32\textwidth]{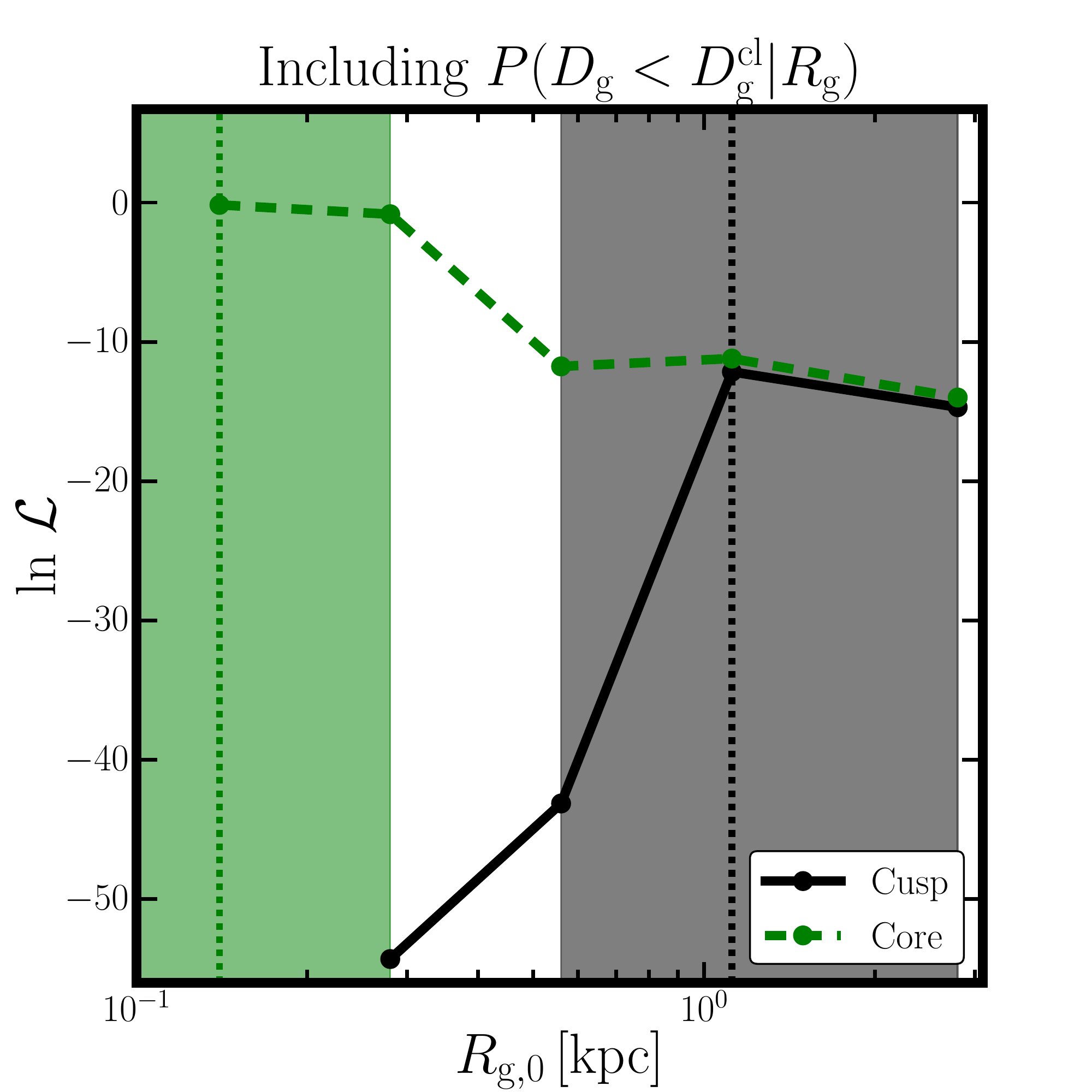}
\includegraphics[width=.32\textwidth]{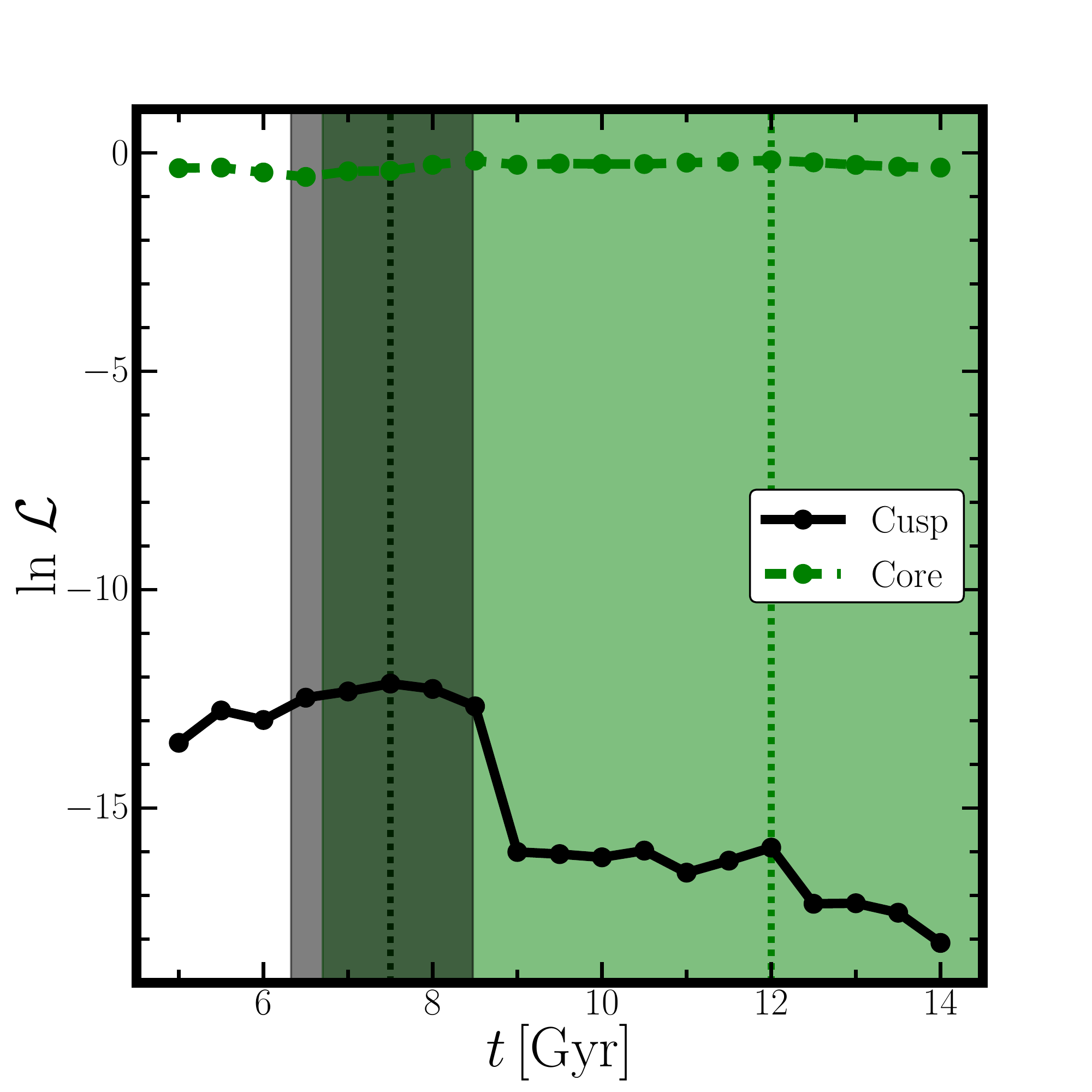}
\caption{
Comparison of the maximum likelihood of the cored and cusped \nbody\ models. The left and middle panels show the maximum likelihood as a function of the initial value of $\Rg$ ($\Rgi$) without (left) and with (middle) $P(\Dg<\Dgcl|\Rg)$. The right panel shows the maximum likelihood as a function of time, $t$, also with $P(\Dg<\Dgcl|\Rg)$. The black solid and green dashed lines are for cusp and core models, respectively. The shaded areas show the 68\% confidence intervals for $\Rgi$ and $t$ for the cored (in green) and cusped (in black) models. The dotted vertical lines show the best-fit values for $\Rgi$ and $t$. Notice that the cored models have a higher likelihood than the cusped models, especially once the probability of observing the cluster at its current projected distance, $P(\Dg<\Dgcl|\Rg)$, is included. As explained in \S~\ref{subsec:DM_profile}, we studied only the clusters that survive for more than $5\,$Gyr.
}
\label{fig:L_profiling}
\end{figure*}

In Fig.~\ref{fig:L_profiling}, we show the maximum likelihood of our models as a function of $\Rgi$, varying all the other parameters ($\Mcl$, $\rh$, and $t$), without (left) and with (middle) $P(\Dg<\Dgcl|\Rg)$ in the likelihood; and as a function of time, $t$ (varying all the other parameters and including $P(\Dg<\Dgcl|\Rg)$, right). The shaded grey and green regions show the 68\% confidence intervals for the parameters $\Rgi$ and $t$ for the cusped (grey) and cored (green) galaxy, respectively. (Assuming that the Wilks' theorem is valid, we used the likelihood ratio to estimate the confidence intervals \citep{Wilks1938}; 
we do not allow our reported  confidence interval to be smaller than the distance between two data points.)

For the clusters in the cored galaxy, the likelihood is bimodal because it is possible to fit the data if the cluster is either in the inner or outer region of the galaxy (see the left panel). For $\Rgi=0.56\,\kpc$, the tidal radius of the cluster is close to its minimum and there it is more difficult to increase the cluster's $\rhl$ up to the observed $13\,\pc$. This leads to the dip in the likelihood at this point. However, including the probability $P(\Dg<\Dgcl|\Rg)$ of observing the cluster at the right position (middle panel) breaks this bimodality, favouring the orbits near the centre with lower $\Rgi$.

For the clusters evolving in the cusped galaxy,
  no star cluster can survive in the inner galaxy $\Rgi < R_{1/2}$ for more than 5\,Gyrs and the likelihood is, therefore, zero for all clusters in that region of parameter space chosen in this study.
 Considering more massive clusters initially does not necessarily lead to a higher probability of survival, because of the increased importance of dynamical friction. Clusters that orbit outside the scale radius ($\Rgi\gtrsim1\,\kpc$) have comparable likelihoods in the cusped and cored models (see left and middle panels) because they are similar by construction at large radii (compare the green and black dashed lines in the right panel of Fig.~\ref{fig:EriII_profile}). (A measurement of the 3D position of the cluster in Eri~II would allow us to completely rule out cusped models. However, to measure a $1\,\kpc$ offset from Eri~II, an accuracy of $0.006\,$mag is needed. Even with RR Lyrae, it is only possible at present to reach an accuracy of $0.05\,$mag.)

The right panel of Fig.~\ref{fig:L_profiling} shows the maximum likelihood at different times. The black line is for clusters in a cusped galaxy, for which the best fits are the models between $6.5$ and $8\,\Gyr$ old. The green dashed line is for the  cored DM profile, for which the best fits are all models with $t\gtrsim7\,\Gyr$. In the cored galaxy, the best fit models are those that survive in the inner part of the galaxy where they can easily expand up to $\sim17\,\pc$ and survive for $14\,\Gyr$. For the cusped galaxy, we can only reproduce the observed properties of Eri~II star cluster for a small amount of time and for a small range of $\Mcli$ and $\rhi$. 
To reproduce Eri~II's star cluster in a cusped galaxy, it must therefore have an age of $6.5 - 8\,\Gyr$. This provides another testable prediction that could fully rule out cusped models. 

Finally, even if we accept a high inclination of the orbital plane and the required orbital phase for the cusped case, its star clusters give a poorer fit to the observations than the cored case, because the clusters are not able to expand enough to match the data. We discuss this in more detail, next.

\subsection{Best-fit star cluster models}\label{sec:best_fit_models}

\begin{figure*}
\center
\includegraphics[width=.48\textwidth]{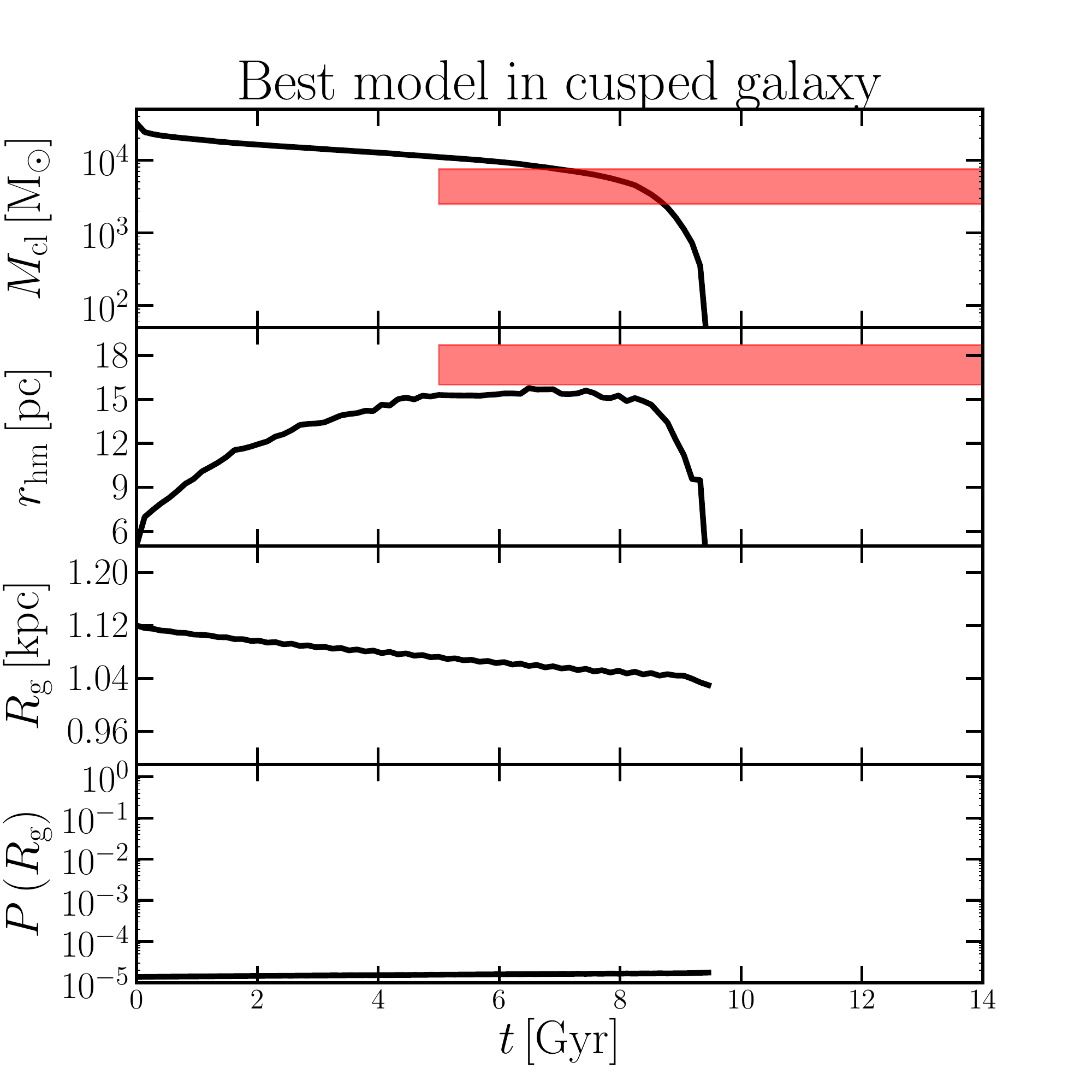}
\includegraphics[width=.48\textwidth]{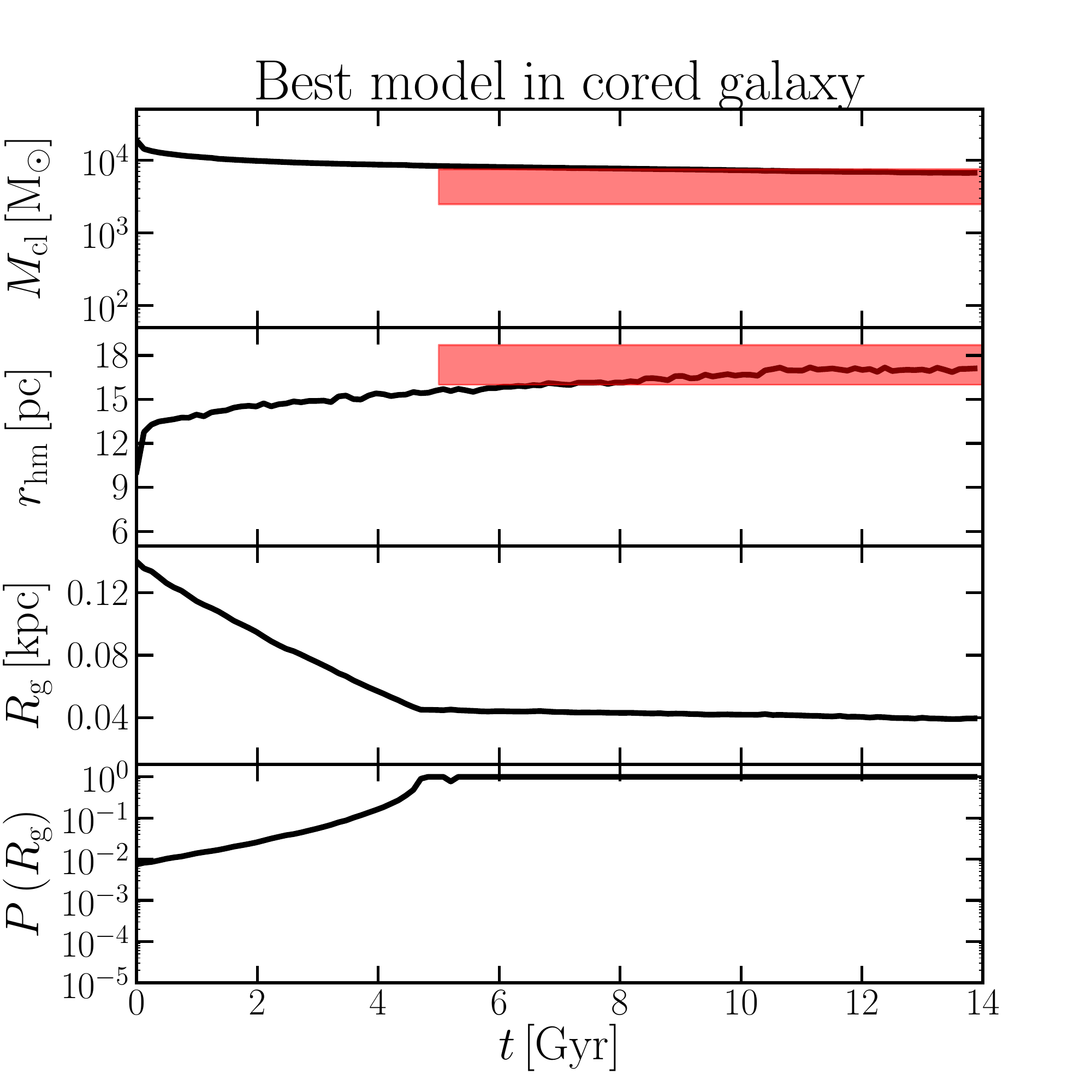}
\caption{(a): Evolution of $\Mcl$, $\rh$, $\Rg$ and the probability to observe a cluster in projection at $\Dg\leq45\pc$, for the best-fit cusped model. Right panel: as left panel but for the best-fit cored model. The red shaded regions show the 68\% confidence intervals of the data. (We only have a lower limit for the age of the cluster, because stars younger than $5\,\Gyr$ have not been observed; see \S\ref{sec:intro}.)}
\label{fig:Nbody}
\end{figure*}

The best-fit model in the cusped galaxy has $\Mcli\sim3.2\times10^4\,\Msun$, $\rhi=5\,\pc$, and $\Rgi=1.12\,\kpc$. The best-fit model in the cored galaxy has $\Mcli\sim1.9\times10^4\,\Msun$, $\rhi=10\,\pc$, and $\Rgi=0.14\,\kpc$. Fig.~\ref{fig:Nbody} shows the evolution of $\Mcl$, $\rh$, $\Rg$ and $P(\Dg<\Dgcl|\Rg)$ for these two models. 
The red shaded areas show the 68\% confidence intervals of the data. (We only have a lower limit for the age of the cluster, because stars younger than $5\,\Gyr$ have not been observed; see \S\ref{sec:intro}.) In  Fig.~\ref{fig:Nbody}, we show that the properties of the star cluster in a cored DM profile reproduce the properties of the observed star cluster for all times $\simgt 5$\,Gyr, whereas in the cusped case the cluster has to be observed at a specific time. Notice, however, that the cusped model is always at tension with the data, with its size, $\rh$, never quite reaching high enough to match Eri~II's star cluster.

As discussed in \S\ref{sec:intro}, we expect the density of the star cluster to reach an equilibrium due to relaxation-driven expansion and the tidal pruning of high-energy escaper stars. In the left panel of Fig.~\ref{fig:Nbody}, we see this process happening between $5$ and $9\,\Gyr$ for the cusped model. Over this period, the cluster evolves at an approximately constant $\rh/\rJ$ \citep{H61}, where $\rJ$ is the `Jacobi' or tidal radius. As a result, the cluster shrinks as $\rh\propto M^{1/3}$ while it loses mass, and it only has a large $\rh$ for a limited time (few Gyr). The cluster in the cored galaxy also expands, but  $\simgt 5$\,Gyr  the star cluster evolves at roughly constant $\Mcl$ and constant $\rh$. This is because the escape rate is very small in compressive tides,  and the cluster evolves towards a near isothermal equilibrium configuration, in which the cluster is in virial equilibrium with the tides (\citealt*{2011MNRAS.414.2728Y}; \citealt{2015MNRAS.447L..40B}; \citealt*{2017MNRAS.468L..92W}). This implies that it is more likely to find a cluster in this phase, because it can be in this quasi-equilibrium configuration for a long time ($\gtrsim 10\,$Gyr). 
The asymptotic value of $\rh$ of the cluster in the cored galaxy is in excellent agreement with the data for Eri~II's star cluster (red shaded region). We note that in our $N$-body model the  cluster density within $\rh$ evolves to approximately the same value as the (uniform) DM density, hence the cluster density is literally probing the DM density. 

\subsection{Predicted cluster number density profiles}

\begin{figure*}
\center
\includegraphics[width=.48\textwidth]{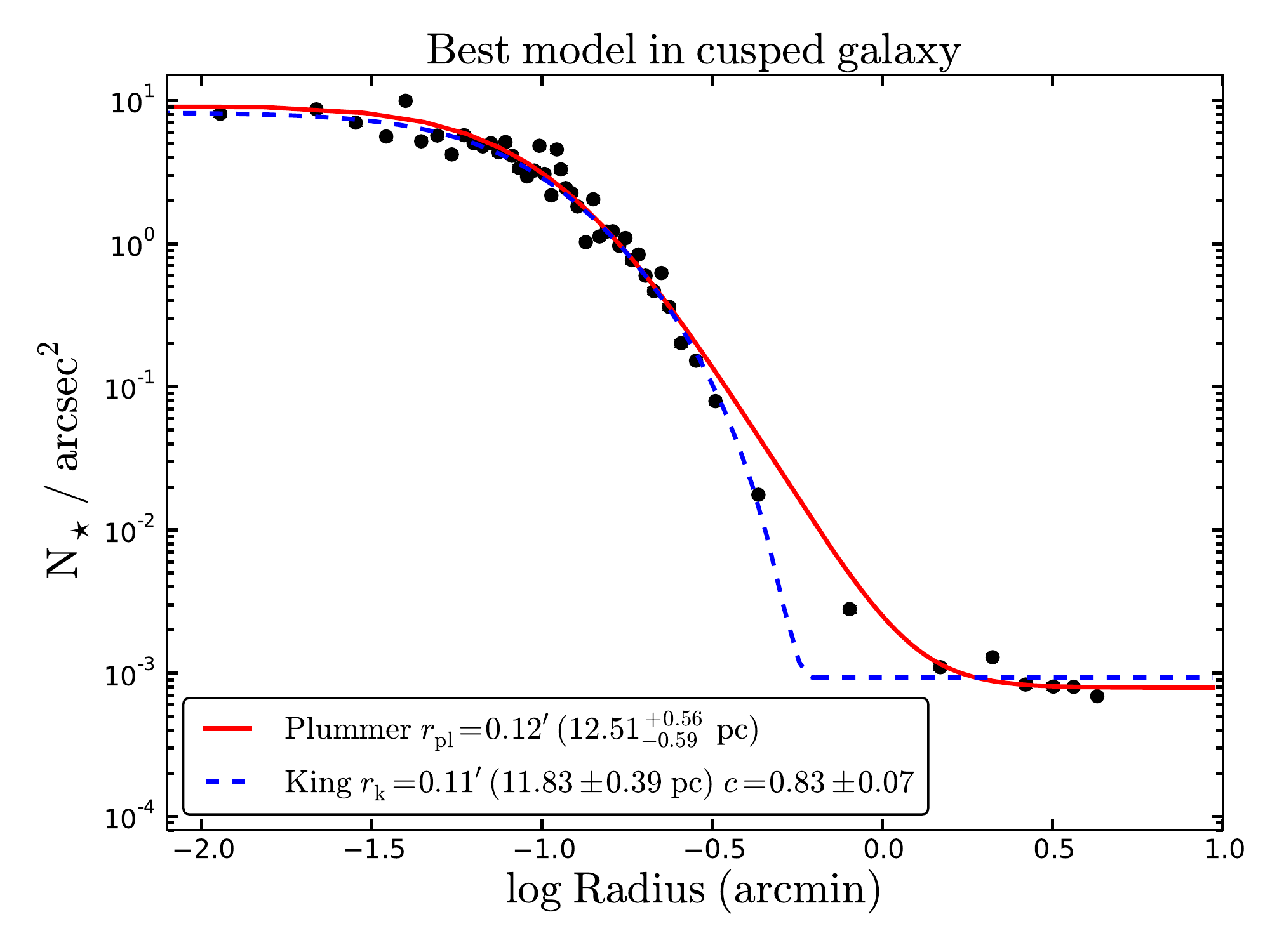}
\includegraphics[width=.48\textwidth]{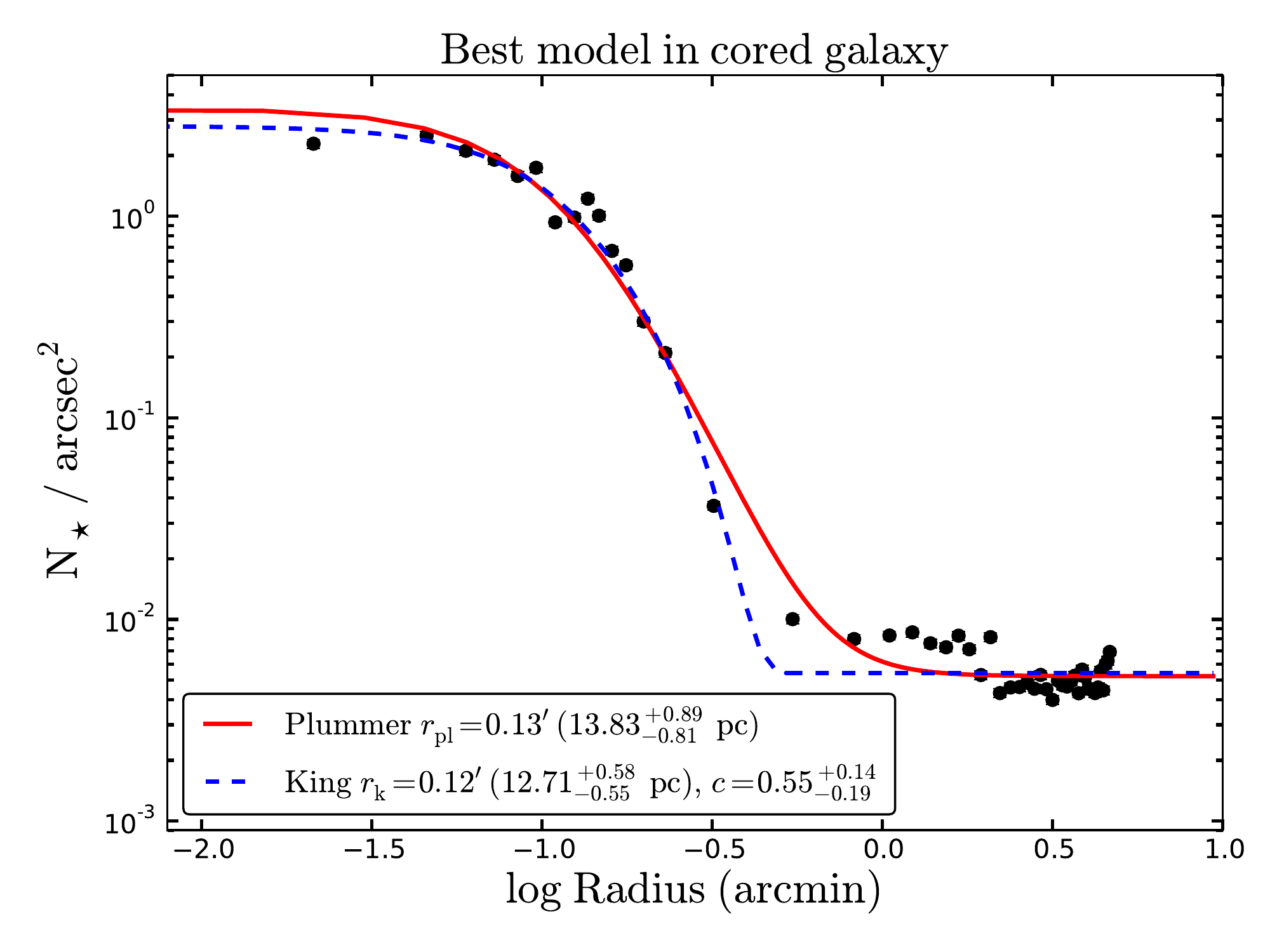}
\caption{(a): Number density profile of the best cusp model at $7.5\,\Gyr$. (b): Number density profile of the best core model at $12\,\Gyr$. 
The solid red and dashed blue lines indicate the best fit Plummer and King profiles, respectively. The number density of background star is estimated using the number density profile of Eri~II \citep{Bechtol2015}. On the right, the number of background stars is higher because the cluster sits in the inner part of Eri~II.
}
\label{fig:sdb}
\end{figure*}

Fig.~\ref{fig:sdb} shows the stellar number density profiles of the best-fit models in the cusped (left) and cored (right) case. \citet{Crnojevic2016} find the structural parameters of Eri~II's star cluster by fitting a Sersic profile to its surface brightness as measured from integrated photometry. It is possible to similarly derive a surface brightness profile from the \nbody\ simulations, however it proved challenging to directly compare the models to the data. Analysing the image from \citep{Crnojevic2016}, we found that the result is very sensitive to the number of bins used, the subtraction of background sources and which bright stars are masked -- all of which can change the result of the fitting. Therefore, for our analysis we used a different approach in which the data from the \nbody\ models are not binned. We used only bright stars (massive stars) that are observable. In our case, we chose only stars that are more massive than $0.75\,\Msun$ \footnote{This mass limit was derived from the observational limit reported in \citet{Crnojevic2016} using {\sc parsec}'s isochrones v1.2S \citep{Bressan2012}.}. Furthermore, we included the background stars using the number density profile of Eri~II reported in \citep{Bechtol2015}, assuming that the stars are uniformly distributed in our simulated field of view.

As can be seen in Fig.~\ref{fig:sdb}, a star cluster that evolves in a DM cusp has a different density profile than a star cluster that evolves in a DM core. In the cored galaxy, clusters have a lower concentration parameter, $c=0.55\pm0.16$, compared to the cluster in the cusped galaxy ($c=0.83\pm0.07$). Here $c\equiv\log(r_{\rm t}/r_0)$, where $r_{\rm t}$ is the truncation radius and $r_0$ is the King/core radius. This means that the core of the cluster is larger (for a given $\rhl$) if it evolves in a DM core. From deeper imaging, it may be possible to derive the projected density profile of the cluster, allowing for a better comparison with our \nbody\ simulations.

\subsection{The effect of varying the mass, concentration and central logarithmic cusp slope of Eri~II's DM halo}\label{subsec:diff_M200_c200}

\begin{figure}
\center
\includegraphics[width=0.5\textwidth]{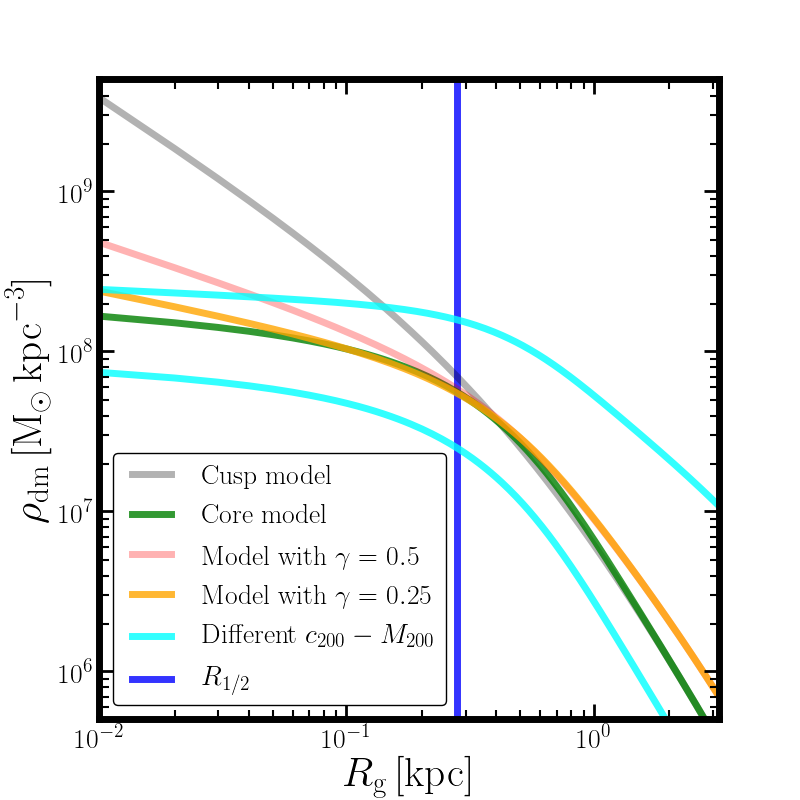}
\caption{The effect of varying the mass, concentration and central logarithmic cusp slope of Eri~II's DM halo. The green and grey lines are the core and cusp models adopt for the simulations (see Figure \ref{fig:EriII_profile}). The cyan lines are the upper and lower envelope of the additional models we ran varying $M_{200}$ and $c_{200}$ and the red and orange lines are the models varying the inner logarithmic cusp slope. Models for which no viable solution for Eri~II's star cluster could be found have shaded lines ($\gamma=0.5$ and $1$).}
\label{fig:different_dens}
\end{figure}

In \S\ref{subsec:DM_profile}, we showed that for dark matter halos that lie on the $M_{200}-c_{200}$ relation, large changes in $M_{200}$ produce only a small change in their inner dark matter density. Here, we test whether we are able to detect such small changes to obtain a constraint on $M_{200}$ from the survival and properties of Eri~II's star cluster alone. To test this, we performed an additional ten simulations varying $M_{200}$ over the range $[1.4\times10^8, 10^{11}]\,\Msun$. (The upper limit of this range is already ruled out by the stellar kinematic measurements for Eri~II \citep{2017ApJ...838....8L}, but serves to test our sensitivity to $M_{200}$.) These models are shown by the cyan lines in Figure \ref{fig:different_dens}.

As anticipated in \S\ref{subsec:DM_profile}, we found that we are not sensitive to even large changes in $M_{200}$ and $c_{200}$ so long as our Eri~II DM halo has a central DM core. By selecting an appropriate $\Mcli$, $\rhi$ and $\Rgi$, we were able in all cases to reproduce the observations within $1\sigma$, similarly to the right panel of Fig.~\ref{fig:Nbody}.

We then tested our sensitivity to the inner logarithmic DM density slope at a fixed DM halo mass of $M_{200}=10^9\,\Msun$, with a concentration set by the $M_{200}-c_{200}$ relation from \citet{2014MNRAS.441.3359D}. We explored two coreNFW models with $n=0.7$ and $n=0.5$, corresponding to Dehnen models (see equation~\ref{eqn:dehnen}) with $(\gamma,M_0,r_0)=(0.25,8.11\times10^8\,\Msun,1.363\,\kpc)$ and $(\gamma,M_0,r_0)=(0.5,9.55\times10^8\,\Msun,1.715\,\kpc)$, respectively. These models are shown by the orange and pink lines in Figure \ref{fig:different_dens}. We found that, with an inner density slope of $\gamma=0.25$, we were still able to find star clusters that survive for longer than $5\,\Gyr$ and reproduce the observations within $2\sigma$, in good agreement with recent results from \citet{2017ApJ...844...64A}. For the galaxy with $\gamma=0.5$, only one cluster (with $\Rgi=0.14\,$kpc, $\rhi=5\,$pc and $\Mcli\sim32,000\,\Msun$) survived for more than $5\,\Gyr$. However, this cluster gave a poor match to Eri~II's star cluster since its $\rh$ expanded up to $12\,$pc and then shrank as the cluster's orbit decayed to the centre of the galaxy. A larger inner density profile slope means a smaller tidal radius for the star cluster and, thus -- all other parameters being equal -- a shorter dissolution time.

\subsection{A nuclear star cluster in Eri~II?}\label{subsec:nuclear_cluster}

Eri~II's star cluster is offset from the photometric centre of Eri~II by $\sim 45$\,pc \citep{Crnojevic2016}. However, given the uncertainties on the photometric centre of Eri~II, we consider here the possibility that Eri~II's star cluster is in fact a nuclear star cluster, defining the centre of the galaxy. A star cluster at the centre of a galactic potential will expand due to two-body relaxation, but without the limiting tidal field. Thus, in a cusped galaxy, a cluster that is tidally destroyed close to the centre of the galaxy (for example, the clusters in the cusped models with $\Rgi=0.14\,$kpc; see \S\ref{sec:best_fit_models}) may still survive if placed at $\Rgi=0.0$\,kpc. As pointed out by \citet{2017ApJ...844...64A}, this could provide a route to Eri~II having a DM cusp without destroying its low density star cluster. To test this, we run six additional simulations of a star cluster with \mbox{$\Mcli\sim$25,000$\,\Msun$} and \mbox{$\rhi = (1, 5,10)\,$pc}, set up to lie at the centre\footnote{We set up these star clusters as Plummer spheres embedded self-consistently in their host Dehnen dark matter halos, using the {\sc mkspherical} program from the {\sc agama} framework \citep[\url{https://github.com/GalacticDynamics-Oxford/Agama}; Vasilev, in prep.; see ][ for another application of th4541241is method]{2017ApJ...848...10V}.} of the cored and cusped Eri~II DM halos described in \S\ref{subsec:DM_profile}. 

\begin{figure}
\center
\includegraphics[width=0.5\textwidth]{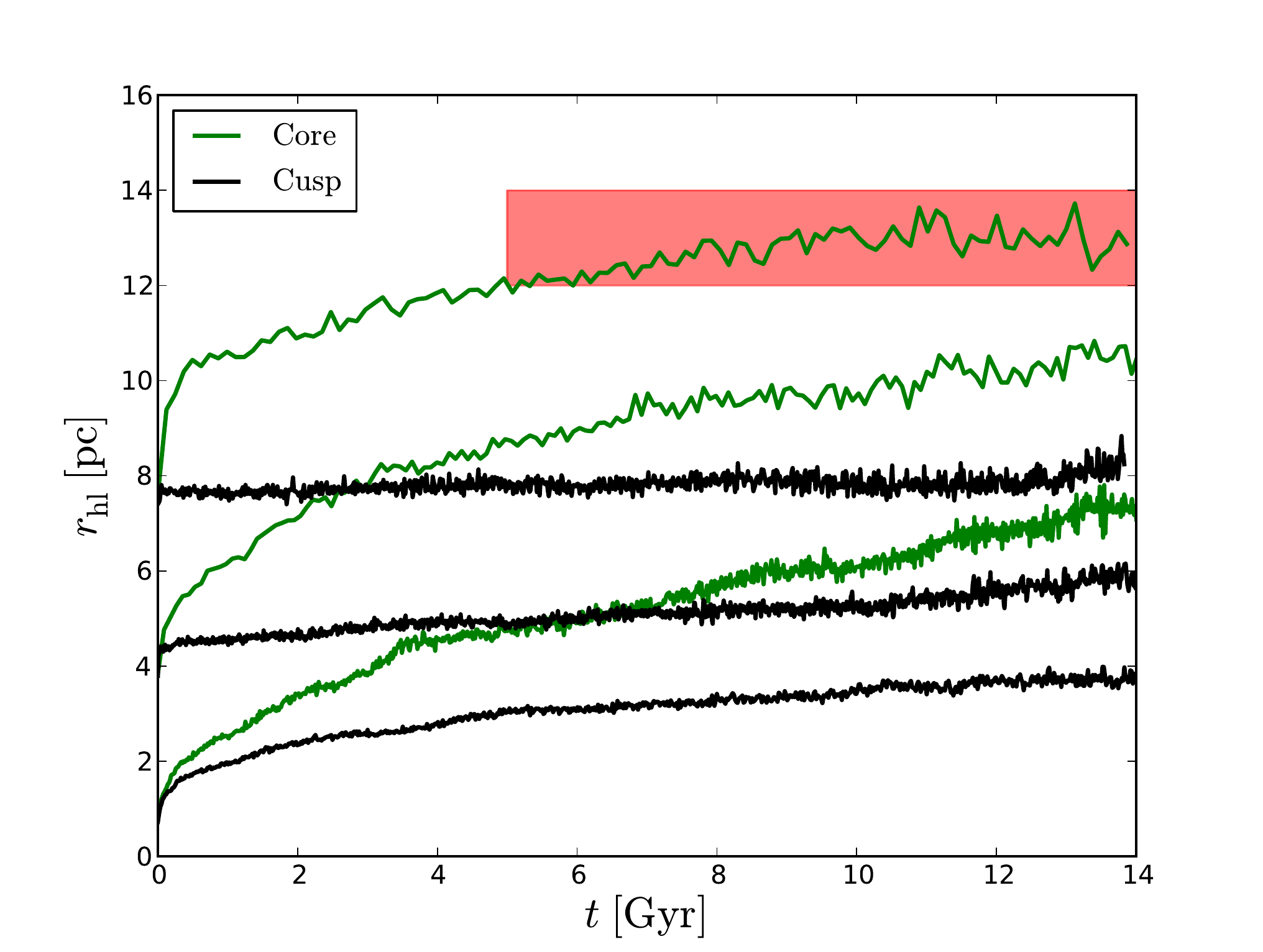}
\caption{Evolution of $\rhl$ for the clusters in the centre of a cusped (black lines) and cored (green lines) galaxy, with $\rhi= (1,5,10)\,$pc. The red shaded region shows the 68\% confidence intervals of the data.}
\label{fig:evo_rhl_centre}
\end{figure}

\begin{figure}
\center
\includegraphics[width=0.5\textwidth]{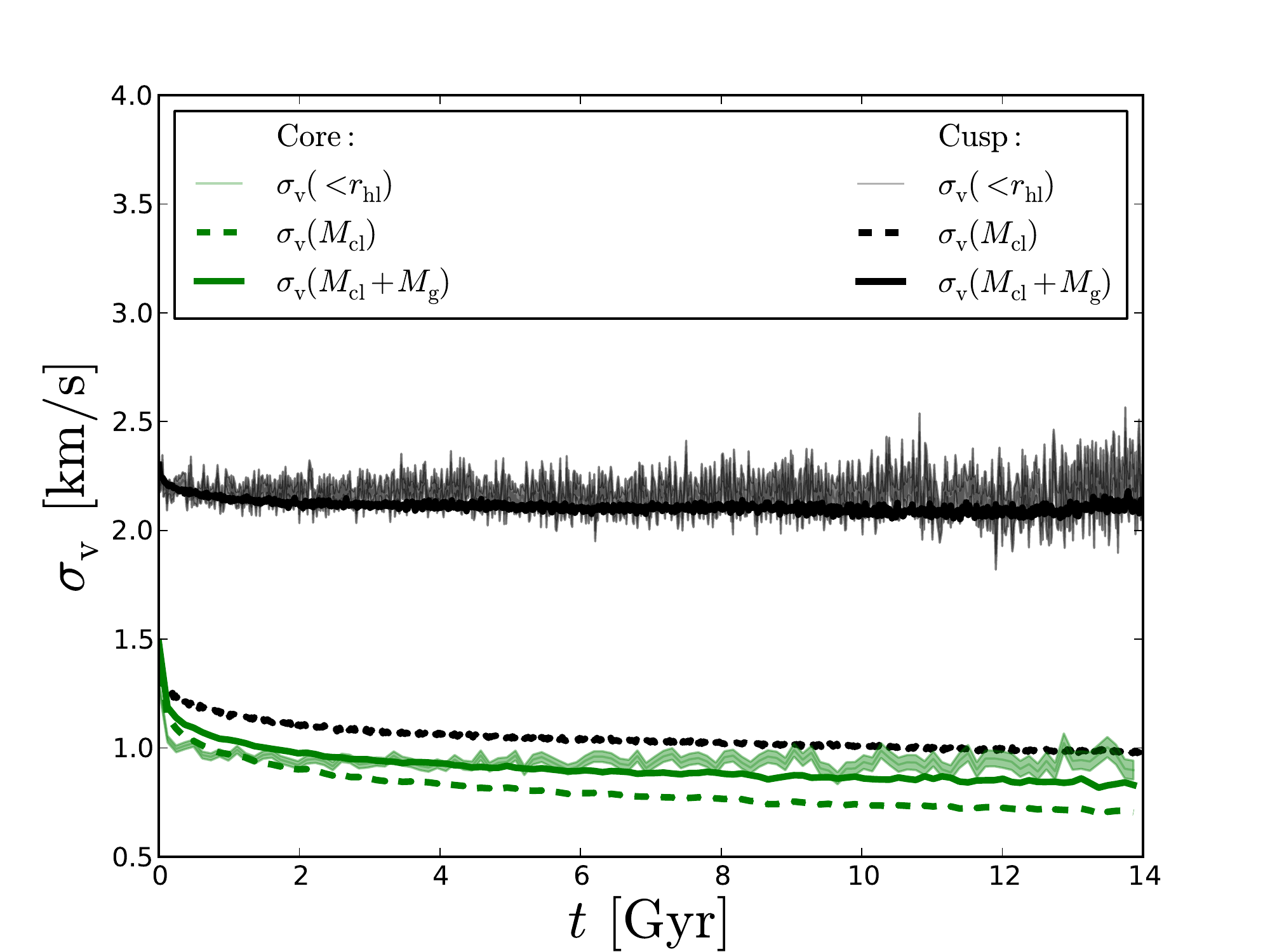}
\caption{Evolution of $\sigv$ for clusters with $\rhi= 10\,$pc, in the centre of a cored (green lines) and cusped (black lines) galaxy. The shaded lines represent the $\sigv$ of the observable stars within $\rhl$, estimated from their line-of-sight velocity. The solid and dashed lines are the inferred $\sigv$ using the `Jeans estimator' from \citet{2009ApJ...704.1274W}. For the dashed lines we used $\Mcl$ within $\rhl$, whereas for the solid lines we use the sum $\Mcl$ and the mass of the galaxy ($M_{\rm g}$) within $\rhl$.}
\label{fig:sig_centre}
\end{figure}

In Fig.~\ref{fig:evo_rhl_centre}, we show the evolution of $\rhl$\footnote{This is derived by multiplying $\rh$ of the observable stars (defined as being only those stars more massive than $0.75\,\Msun$ and excluding any dark remnants) by $3/4$ to correct for projection effects.} for these simulations in a cored (green lines) and a cusped (black lines) DM halo. In the cusped case, for $\rhi=10\,\pc$, $\rhl$ does not expand because the cluster is mainly dark matter dominated and so the cluster stars trace the underlying dark matter potential. If Eri~II's star cluster formed in the centre of Eri~II, it must have formed with a size similar to that observed today, but with almost double its current mass\footnote{A cluster in the centre of a cusped galaxy loses mass mainly due to stellar evolution.}.
In Fig.~\ref{fig:sig_centre}, we show the velocity dispersion $\sigv$, estimated for observable stars within $\rhl$, for the simulations with $\rhi=10\,\pc$ in a cusped (black lines) and cored (green lines) galaxy. In the \nbody\ simulations, both $\Mcl$ and the mass of the galaxy ($M_{\rm g}$) within $\rhl$ are known. Thus, using the `Jeans estimator' formula from \citet{2009ApJ...704.1274W}, we can estimate the $\sigv$ due to $\Mcl$ ($\sigv(\Mcl)$; dashed lines), and due to the combined mass of the cluster and the galaxy ($\sigv(\Mcl+M_{\rm g})$; solid lines). From this, we conclude that if Eri~II has a DM cusp and hosts a nuclear star cluster, then its star cluster will be DM dominated, with a velocity dispersion\footnote{Note that none of the clusters in our cusped simulations have a $\rhl$ large enough to be consistent with observations (see Fig. \ref{fig:evo_rhl_centre}). However, in a DM cusp, the enclosed mass goes as $M \propto r^2$ and so $\sigma_v^2 \propto r$. Thus, increasing the size of the cluster will increase $\sigma_v$, which is why our results provide a lower bound on the dispersion.} $\sigv>2.5\,\kms$. By contrast, if Eri~II has a DM core, its star cluster will have a much lower dispersion of $\sigv<1.0\,\kms$. These results are in good agreement with \citet{2017ApJ...844...64A}.

Although we have found viable models where Eri~II's star cluster sits at the very centre of a dense DM cusp, we emphasise that -- even without a measurement of $\sigv$ -- we disfavour these. Firstly, Eri~II's star cluster is observed to be clearly offset from the photometric light centre of Eri~II. Secondly, our cored model in \S\ref{sec:cuspVScore} naturally reproduces Eri~II's star cluster properties without any fine tuning. If Eri~II's star cluster sits at the centre of a dense DM cusp, then its properties must be set by its birth properties, requiring more fine tuning than for the cored case.

\section{Discussion}\label{sec:discussion}
Our key result is that we favour a DM core over a cusp in the ultra-faint dwarf galaxy Eri~II. In models with a DM cusp, Eri~II's star cluster is rapidly destroyed by tides, whereas in cored models the star cluster survives for more than a Hubble time, naturally reaching an asymptotic size and mass consistent with observations. We found that this occurs for logarithmic DM cusp slopes shallower than $\gamma = 0.25$, where $\gamma$ is the central exponent in the Dehnen profile (equation \ref{eqn:dehnen}), independently of large changes in the assumed DM halo mass or concentration. The only hope for retaining a DM cusp in Eri~II is if its star cluster lies at the very centre of the cusp. We found that such a model can work, but is disfavoured by the observed offset between Eri~II's photometric light peak and the projected position of its star cluster. Such a model could be completely ruled out if the velocity dispersion of Eri~II's star cluster is observed to be $\sigv<1.0\,\kms$. These results are in excellent agreement with a recent study by \citet{2017ApJ...844...64A}.

Our mass model for Eri~II has a DM core size set by the projected half light radius of the stars $R_{1/2} \sim 0.28$\,kpc (see Fig.~\ref{fig:EriII_profile}). However, the data only require that there is a dark matter core where we see Eri~II's star cluster today, at a projected distance of 45\,pc from the photometric centre of Eri~II. Dynamical friction stalling occurs when the tidal radius of the star cluster approximately matches its galactocentric distance \citep{Read2006a,Goerdt2010,Petts2015,Petts2016}. From this, we can derive a minimum DM core size for Eri~II of $r_{\rm c} > 45$\,pc. In this section, we explore what such a dark matter core means for galaxy formation and the nature of DM.

\subsection{Dark matter heating}

The size and density of the DM core we find in Eri~II is in excellent agreement with predictions from the `DM heating' model of \citetalias{Read2016a}. However, in the \citetalias{Read2016a} model such a complete DM core would require several Gyrs of star formation that may be inconsistent with Eri~II's stellar population (see \S\ref{sec:intro}). However, core formation can be made more efficient if it occurs at high redshift when Eri~II's DM halo was less massive \citep{2014ApJ...789L..17M}, or if it owes primarily to angular momentum transfer from cold gas clumps sinking by dynamical friction to the centre of the dwarf \citep{2001ApJ...560..636E,2015MNRAS.446.1820N}. Given these complications, following \citet{2012ApJ...759L..42P}, \citetalias{Read2016a} and \citet{Read2017}, we focus here on the energy required to unbind Eri~II's dark matter cusp:

\begin{equation}
\frac{\Delta E}{\Delta W} = \frac{M_*}{\langle m_*\rangle \Delta W}\xi\epsilon_{\rm DM}
\end{equation}
where $\Delta E$ is the total integrated supernova energy, $\Delta W$ is the energy required to unbind the dark matter cusp, $M_*$ is the stellar mass, $\langle m_*\rangle = 0.83$ is the mean stellar mass, $\xi = 0.00978$
is the fraction of mass in stars that go supernova (i.e. those with mass $m_* > 8\,\Msun$), and $\epsilon_{\rm DM} = 0.0025$ is the coupling efficiency of the SNe energy to the dark matter. We assume a Chabrier initial stellar mass function over the stellar mass range $0.1 < m_*/\Msun < 100$ \citep{2003ApJ...586L.133C}. We assume a coreNFW profile when calculating $\Delta W$ and we take $\epsilon_{\rm DM}$ from the simulations in \citetalias{Read2016a}. As such, our results are only useful in assessing, at an order-of-magnitude level, whether there is sufficient supernova energy in Eri~II's stellar population to form its apparent central DM core \citep[e.g.][]{2015ApJ...806..229M}.

\begin{figure}
\center
\includegraphics[width=0.49\textwidth]{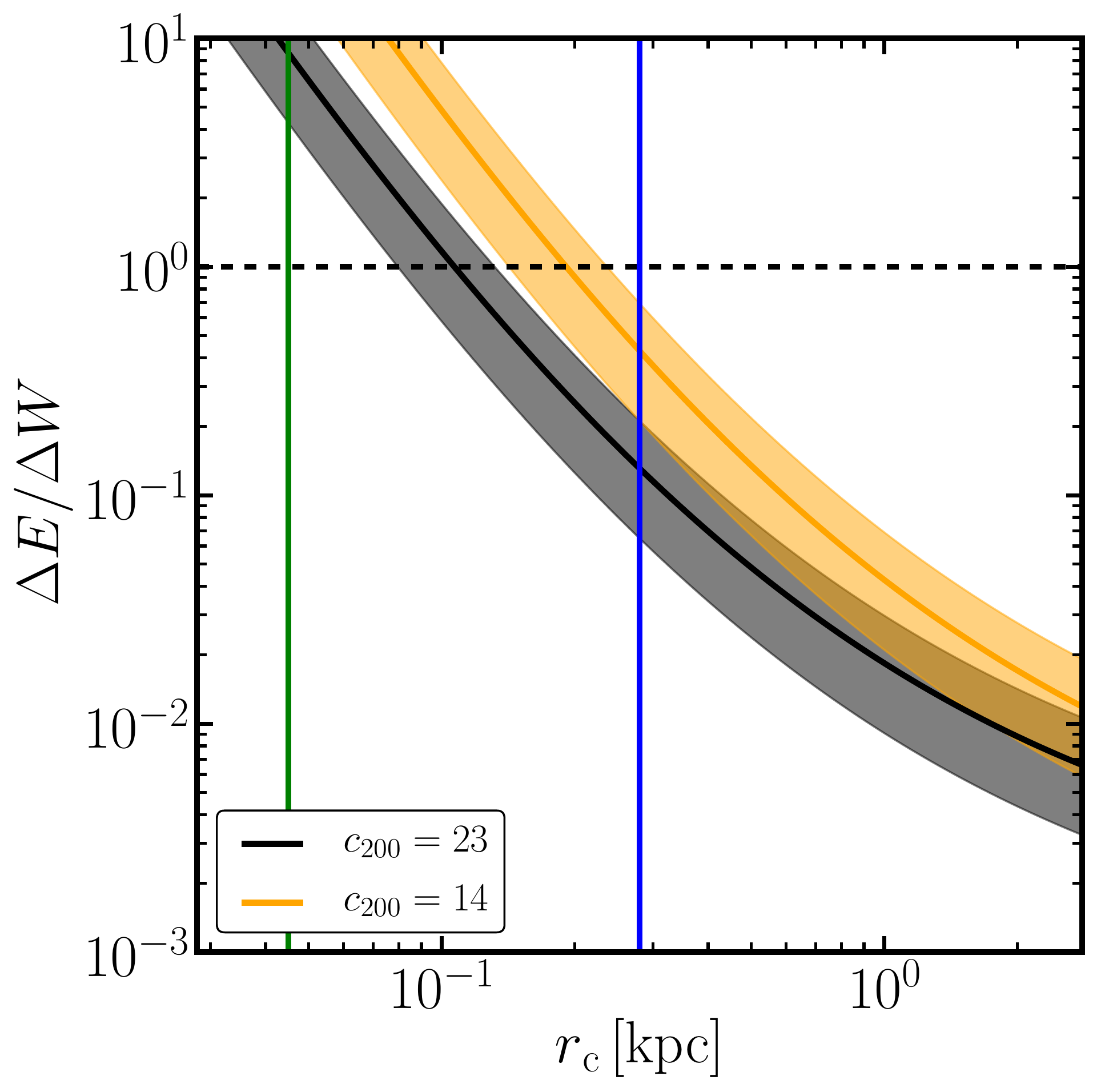}
\caption{The energy required to unbind Eri~II's DM cusp as a function of the DM core size, $r_{\rm c}$. We assume for this plot a Chabrier IMF \citep{2003ApJ...586L.133C} and a coupling efficiency between supernovae and dark matter of $\epsilon_{\rm DM} = 0.25\%$ (\citetalias{Read2016a}). The horizontal dashed line marks $\Delta E / \Delta W = 1$. Above this line, there is enough integrated supernova energy to unbind the cusp; below there is insufficient energy. The vertical green line marks the minimum core size, $r_{\rm c,min} = 45$\,pc set by the current projected position of Eri~II's star cluster. The vertical blue line marks the core size assumed in this work, $r_{\rm c} = R_{1/2} = 0.28$\,kpc. The black and orange shaded regions assume a DM halo mass of $M_{200} = 5 \times 10^8\,\Msun$ with a stellar mass $M_* = 8.3^{+5.1}_{-4.2} \times 10^4\,\Msun$ \citep[][assuming three times their uncertainties]{Bechtol2015}, and a concentration parameter of $c_{200} = 23$ and $c_{200} = 14$, respectively. These are the upper and lower 68\% confidence intervals of $c_{200}$ in $\Lambda$CDM \citep{2014MNRAS.441.3359D}.
}
\label{fig:eridanus_energy}
\end{figure}

In Fig.~\ref{fig:eridanus_energy}, we plot $\Delta E / \Delta W$ as a function of the DM core size $r_{\rm c}$. The horizontal dashed line marks $\Delta E / \Delta W = 1$. Above this line, there is enough integrated supernova energy to unbind the cusp; below there is insufficient energy. The vertical green line marks the minimum core size, $r_{\rm c,min} = 45$\,pc set by the current projected position of Eri~II's star cluster. The vertical blue line marks the core size assumed in this work, $r_{\rm c} = R_{1/2} = 0.28$\,kpc. The black and orange shaded regions assume a DM halo mass of $M_{200} = 5 \times 10^8\,\Msun$ with a stellar mass $M_* = 8.3^{+5.1}_{-4.2} \times 10^4\,\Msun$ \citep[][assuming three times their uncertainties]{Bechtol2015}, and a concentration parameter of $c_{200} = 23$ and $c_{200} = 14$, respectively. These are the upper and lower 68\% confidence intervals of $c_{200}$ in $\Lambda$CDM \citep{2014MNRAS.441.3359D}. (Note that the \citetalias{Read2016a} simulations assume the upper envelope of this $c_{200}$ range and so make core formation maximally difficult.)

As can be seen in Fig.~\ref{fig:eridanus_energy}, there is plenty of energy to produce the minimum core size $r_{\rm c} = 45$\,pc independently of the assumed $c_{200}$ or $M_*$. However, for the assumed DM halo mass and supernova energy coupling efficiency that we assume here, a low $c_{200}$ and high $M_*$ for Eri~II are required to produce a core as large as $r_{\rm c} = R_{1/2} = 0.28$\,kpc.

Although the \citetalias{Read2016a} models are able to produce a DM core in Eri~II, most other simulations in the literature to date do not find DM cores in halos below $M_{200} \sim 7.5 \times 10^{9}\,\Msun$ (e.g. \citealt{2015MNRAS.454.2092O,2015MNRAS.454.2981C,2016MNRAS.456.3542T}, but see \citealt{2014ApJ...789L..17M}). We can understand the origin of this discrepancy from Fig.~\ref{fig:ms_mh}. As can be seen, the \citetalias{Read2016a} simulations (magenta squares) produce a similar total stellar mass to the \citet{2015MNRAS.454.2981C} (yellow squares) and \citet{2015MNRAS.454...83W} (green squares) simulations but in halos an order of magnitude lower in mass. (Note that the \citet{2015MNRAS.454...83W} simulations are the same as those discussed in \citet{2016MNRAS.456.3542T}.) This is why cusp-core transformations in the \citetalias{Read2016a} simulations are energetically feasible. The simulations that find no cores below $M_{200} \sim 7.5 \times 10^{9}\,\Msun$ form almost no stars below this mass scale and so do not have enough integrated supernova energy to unbind the dark matter cusp (see also the discussion in \citealt{Read2017}). Understanding this apparent discrepancy between the data and simulations in Fig.~\ref{fig:ms_mh}, and the differences between numerical models, remains an open and important problem.

If Eri~II is found to have a purely old stellar population, with too few supernovae to provide the energy required to unbind its cusp, then we may be forced to move to models beyond $\Lambda$CDM. We consider some of these, next.

\subsection{Beyond $\Lambda$CDM}

The small scale puzzles in $\Lambda$CDM (see \S\ref{sec:intro}) have motivated the community to consider alternative models. Some of these can solve the cusp-core problem without recourse to baryonic `DM heating'. The most popular of these to date is Self-Interacting Dark Matter (SIDM; \citealt{2000PhRvL..84.3760S}). The latest SIDM models have a velocity-dependent interaction cross section $\sigma/m(v/v_m)$, where $\sigma$ is the dark matter interaction cross section, $m$ is the mass of the dark matter particle and $v_m$ is a velocity scale \citep[e.g.][]{2016PhRvL.116d1302K,2017MNRAS.470.1542S}. This is required for the models to be consistent with constraints from weak lensing on galaxy cluster scales that favour $\sigma/m < 0.5\,{\rm cm}^2/{\rm g}$ \citep[e.g.][]{Harvey2015}, while maintaining a much higher $\sigma/m \sim 2-3\,{\rm cm}^2/{\rm g}$ required to produce large DM cores in nearby gas rich dwarf galaxies \citep[e.g.][]{2016PhRvL.116d1302K}.

Since our focus here is on one low-mass dwarf, we consider instead a simple {\it velocity independent} SIDM model. Following \citet{2017MNRAS.470.1542S}, the velocity independent interaction cross section can be written as:

\begin{equation}
\frac{\sigma}{m} = \frac{\sqrt{\pi}\,\Gamma}{4 \rho_{\rm NFW}(r_{\rm c}) \sigma_v(r_{\rm c})}
\end{equation}
where $r_{\rm c}$ is the coreNFW DM core size (\citetalias{Read2016a}), $\rho_{\rm NFW}$ is the {\it initial} DM density at $r_{\rm c}$, $\Gamma = 0.4$\,Gyr$^{-1}$ is the SIDM interaction rate (taken from coreNFW fits to  numerical simulations in SIDM; \citealt{2017MNRAS.470.1542S}) and: 

\begin{equation}
\sigma_v(r_{\rm c})^2 = \frac{G}{\rho_{\rm NFW}}\int_{r_{\rm c}}^\infty \frac{M_{\rm NFW}(r')\rho_{\rm NFW}(r')}{r'^2}dr'
\end{equation}
is the velocity dispersion of the DM at $r_{\rm c}$ (assuming that this model is isotropic).
Using $r_{\rm c} > 45$\,pc, we find $\sigma/m > 0.24\,{\rm cm}^2/{\rm g}$ which is consistent with all known SIDM constraints to date.

Another model that could explain Eri~II is ultra-light axions (e.g. \citet{Gonzales2016} and references therein). Assuming an ultra-light axion mass of $m_a \sim 10^{-22}\,{\rm eV}$, we obtain for our Eri~II halo model, $r_{\rm c} \sim 0.6 \,\kpc$ which is consistent with our minimum core size $r_{\rm c} > 45$\,pc.

Therefore, alternative DM models such as SIDM and \mbox{ultra-light} axions can produce a core in the ultra-faint Eri~II dwarf galaxy.

\subsection{Implications of the initial cluster properties}\label{subsec:implication_cluster_prop}

The initial mass of the star cluster in the cored galaxy ($\Mcli\simeq1.9\times10^4\,\msun$) suggests that this star cluster resembled a young massive cluster, similar to those we see in the disc of the Milky Way (e.g. Arches, Westerlund 1, NGC\,3603). The initial radius  ($\rhi\simeq10\,\pc$) is relatively large compared to these low redshift analogues, which have typical radii of a few pc \citep{2010ARA&A..48..431P}. We note that the model with $\rhi\simeq5\,$pc expands up to  $\rh\sim15\,\pc$ and fits the data reasonably well. Our models did not include primordial mass segregation in the cluster initial conditions. If we had assumed that the massive stars formed more towards the centre of the cluster, it would have expanded more as a result of stellar mass loss \citep[e.g.][]{2017MNRAS.467..758Z}, allowing for more compact initial conditions. The present day mass of stars and stellar remnants that once belonged to the cluster is $\sim10^4\,\msun$, or $\sim12\%$ of the total stellar mass in Eri~II. Such a high cluster formation efficiency has been reported in other dwarf galaxies when considering metal-poor stars only \citep{2014A&A...565A..98L}.

\subsection{Comparison with other work in the literature}
Concurrent with our work, \citet{2017ApJ...844...64A} have recently modelled Eri~II's central star cluster sinking under dynamical friction, finding that it cannot survive long in a cusped potential, in good agreement with our results here (They also report similar results for the star cluster in Andromeda XXV.) The key difference between our studies is that we model the internal structure of Eri~II's star cluster, accounting for two-body relaxation and stellar evolution, for the first time. \citet{2017ApJ...844...64A} used a collisionless $N$-body code which cannot capture two-body relaxation effects that drive the expansion of the cluster over time\footnote{\citet{2017ApJ...844...64A} argue that collisional effects can be ignored in extended and faint star clusters. However, this depends on the assumed initial conditions for the cluster. The initial half-mass relaxation time \citep{Spitzer1987} of our best-fit cluster model is $\sim2.2\,$Gyr, computed assuming $\ln\Lambda=\ln(0.02N)$ \citep{1996MNRAS.279.1037G}. Since the cluster is older than this, two-body relaxation was important during the evolution of this cluster.}. The key advantage of modelling the collisional effects in Eri~II's star cluster is that its final mass and size then depend on the local tidal field and, therefore, on the mass distribution at the centre of Eri~II. This gives us an additional probe of the central dark matter density profile in Eri~II that goes beyond a survival argument. 

\section{Conclusions}\label{sec:conclusions}

We have presented a new method for probing the central dark matter density in dwarf galaxies using star clusters. Low-mass star clusters orbiting in the tidal field of a larger host galaxy are expected to reach an equilibrium size due to relaxation-driven expansion and the tidal pruning of high-energy escaper stars. We have used the \nbsixdf\ collisional $N$-body code, which includes stellar evolution and dynamical friction, to show that this is indeed the case. As a first application, we have applied our method to the recently discovered ultra-faint dwarf, Eri~II. This has a lone star cluster that lies some $\sim 45$\,pc from its centre in projection. Using a suite of 226 full $N$-body simulations of the star cluster, we showed that models with a central dark matter core (with an inner logarithmic density slope of $\gamma < 0.25$) are favoured over those with a dark matter cusp. A DM core naturally reproduces the size 
and the projected position of Eri~II's star cluster. By contrast, dense cusped galaxy models require the cluster to lie implausibly far from the centre of Eri~II \mbox{($>1\,\kpc$)}, with a high inclination orbit ($i>87.43^\circ$) that must be observed at a special orbital phase (<3 per cent of the orbital period). 

Our models make several clear predictions that can be tested with deeper observations. If Eri~II is cored, then: 
\begin{itemize}
\item the cluster can have any age older than $\sim7\,\Gyr$ (as compared to a narrow age range of $6.5-8\,$Gyr in the cusped case);
\item there are no tidal tails associated with the cluster;
\item the cluster has a low concentration ($c\sim0.5$ as compared to $c\sim0.8$ in the cusped case).
\end{itemize}

We also considered the possibility that Eri~II's star cluster lies at the very centre of a DM cusp, allowing it to survive tidal disruption. This is already disfavoured by the observed offset between Eri~II's photometric light peak and the projected position of its star cluster. However, such a model could be completely ruled out if the velocity dispersion of Eri~II's star cluster is found to be $\sigv<1.0\,\kms$.

We have shown that extended faint star clusters can survive at the centres of dwarf galaxies with DM cores. Such faint star clusters could be liberated from their host dwarf galaxy by Galactic tides that act more efficiently on cored dwarfs \citep[e.g.][]{2006MNRAS.367..387R,2010MNRAS.406.1290P}, providing an explanation for some of the recently discovered ultra-faint objects found in the Milky Way \citep[for a discussion see][]{2017MNRAS.466.1741C}.

The presence of a DM core in the ultra-faint dwarf galaxy Eri~II implies that either its CDM cusp was `heated up' by bursty star formation, or we are seeing an evidence for physics beyond CDM.

\section*{Acknowledgments}\label{acknowledgments}
Support for this work was provided by the European Research Council (ERC-StG-335936, CLUSTERS). MG acknowledges financial support from the Royal Society in the form of a University Research Fellowship (URF) and an equipment grant used for the GPU cluster in Surrey. JIR would like to acknowledge support from STFC consolidated grant ST/M000990/1 and the MERAC foundation. We thank Josh D. Simon for the insightful discussions about the interpretation of the HST photometry, Denija Crnojevic for the image of Eri~II, and the referee for comments and suggestions. We are grateful to Sverre Aarseth and Keigo Nitadori for making {\sc nbody6} publicly available, and to Dan Foreman-Mackey for providing the {\sc emcee} software and for maintaining the online documentation; we also thank Mr David Munro of the University of Surrey for hardware and software support. The analyses done for this paper made use of {\sc scipy} \citep{scipy}, {\sc numpy} \citep{numpy} and {\sc matplotlib} \citep{matplotlib}.

\bibliography{EridanusII}
\bibliographystyle{mn2e}

\end{document}